\documentclass[12pt]{iopartJP}
\makeatletter
\renewcommand\@appendixstar{\@@par
 \ifnumbysec 
 \@addtoreset{table}{section}
 \@addtoreset{figure}{section}\fi
 \setcounter{section}{0}
 \setcounter{subsection}{0}
 \setcounter{subsubsection}{0}
 \setcounter{equation}{0}
 \setcounter{figure}{0}
 \setcounter{table}{0}
 \def\thesection{\Alph{section}} 
 \def\theequation{\ifnumbysec
      \Alph{section}.\arabic{equation}\else
      \Alph{section}\arabic{equation}\fi}
 \def\thetable{\ifnumbysec
      \Alph{section}\arabic{table}\else
      A\arabic{table}\fi}
 \def\thefigure{\ifnumbysec
      \Alph{section}\arabic{figure}\else
      A\arabic{figure}\fi}}
\makeatother
\usepackage{amsmath}  
\usepackage{amssymb}
\usepackage{graphicx} 

\usepackage{pdfsync}
\usepackage{tikz}

\usepackage{color}
\usepackage{xcolor}
\usepackage[breaklinks,colorlinks,backref]{hyperref}
\hypersetup{
    colorlinks, 
    linktoc=all, 
    linkcolor=red,  
  citecolor=hyptxt,
  urlcolor=blue}
  \hypersetup{
  citebordercolor=,
  filebordercolor=green,
  linkbordercolor=blue
}
\definecolor{hyptxt}{rgb}{0.7, 0.4, 0.9}
\usepackage[margin=1in]{geometry}

\def\calH{{\cal H }}

\def\H{\mathbb{H}}
\def\R{\mathbb{R}}

\def\C{\mathbb{C}}

\def\ii{\mathrm{i}}

\def\ud{\mathrm{d}}
\def\rg{\rangle}
\def\lg{\langle}

\def\sfC{\mathsf{C}}

\def\mcE{\mathcal{E}}
\def\upa{| \uparrow\,\rg}
\def\dwa{| \downarrow\,\rg}
\def\pua{\lg \uparrow|}
\def\wda{\lg \downarrow|}
\def\Da{\mbox{\Large \textit{a}}}

\numberwithin{equation}{section}

\begin{document}

\title[Baby quantum]{A  baby Majorana quantum formalism}

\author{Herv\'e Bergeron}
\address{Univ Paris-Sud, ISMO, UMR 8214, 91405 Orsay, France}
\ead{herve.bergeron@u-psud.fr}

\author{Evaldo M. F. Curado}
\address{Centro Brasileiro de Pesquisas Fisicas 
and  Instituto Nacional de Ci\^encia e Tecnologia - Sistemas Complexos\\
Rua Xavier Sigaud 150, 22290-180 - Rio de Janeiro, RJ, Brazil}
\ead{evaldo@cbpf.br}

\author{Jean-Pierre  Gazeau}
\address{APC, UMR 7164, Univ Paris  Diderot, Sorbonne Paris Cit\'e,   
75205 Paris, France}
\ead{gazeau@apc.in2p3.fr}

\author{ Ligia M.C.S. Rodrigues}
\address{Centro Brasileiro de Pesquisas Fisicas 
Rua Xavier Sigaud 150, 22290-180 - Rio de Janeiro, RJ, Brazil}
\ead{ligia@cbpf.br}

\vspace{10pt}
\begin{indented}
\item[01]\ June 2017
\end{indented}

%
%
%
%
%
\newpage

\tableofcontents

\newpage


\begin{abstract}
The aim of the present paper is to introduce and to discuss the most basic fundamental concepts of  quantum physics by means of a simple and pedagogical example. An appreciable part of its content presents original results. We start with the Euclidean plane which is certainly a paradigmatic example of a Hilbert space. The pure states form the unit circle (actually a half of it), the mixed states form the unit disk (actually a half of it), and rotations in the plane rule time evolution through  Majorana-like equations involving only real quantities for closed and open systems. The set of pure states or a set of mixed states  solve the identity and they are used for understanding the concept of integral quantization of functions on the unit circle and to give a semi-classical portrait of quantum observables. Interesting probabilistic aspects  are developed.   Since the tensor product of two planes, their direct sum,  their cartesian product, are isomorphic ($2$ is the unique solution to $x^x= x\times x = x+x$), and they are also isomorphic to $\C^2$, and to the quaternion field $\H$ (as a vector space), we describe an interesting relation between  entanglement of real states, one-half spin cat states, and unit-norm quaternions which form  the group SU$(2)$. We explain the most general form of the Hamiltonian in the real plane by considering the integral quantization of a magnetic-like interaction potential viewed as a classical observable on the unit $2$-sphere. Finally, we present an example of quantum measurement with  pointer states lying also in the Euclidean plane. 
\end{abstract}


\section{Introduction} 
\label{intro}

Quantum formalism contains many non-intuitive concepts like state or ray in Hilbert space, superposition, operator and spectrum, entanglement, measurements and collapse etc.  In this paper we intend to introduce these concepts in a  simple way with the objective of making them accessible to  non-specialists. Notwithstanding, in spite of their simplicity, some of our results are original.  

Hilbert spaces in  quantum physics are usually vector spaces on complex numbers, and many of them are infinite-dimensional.  Here our approach is utterly minimalist, 
not only in order to present the quantum formalism in a simple way but also for pedagogical purposes. It is minimalist in the sense that our departure point is the simplest non-trivial Hilbert space, namely the so familiar Euclidean plane \cite{euclid300BC}, denoted  $\R^2$. 
The idea of implementing quantum formalism with real instead of complex Hilbert spaces is  not novel, of course, and can be traced back to Stueckelberg \cite{stueck60} for instance. However, we show that the formalism developed in our article applies as well to unexpected situations, far beyond the mathematical modelling of the microscopical world. 

\begin{figure}
\begin{center}
\includegraphics[width=5in]{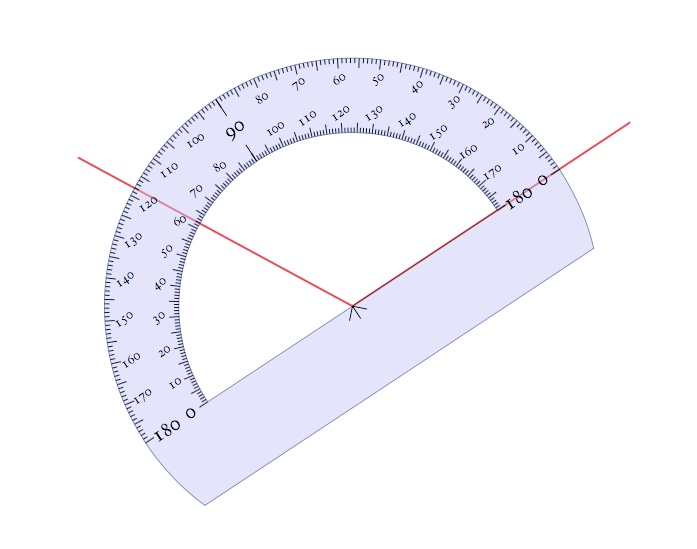}
\caption{Protractor viewed as set of unit rays in the Euclidean plane. The identification with $\mathbb{S}^1/\{-1,1\}$ is clearly apparent on the picture, with $0^{\circ}\equiv -180^{\circ}$, $10^{\circ}\equiv -170^{\circ}$, etc, sign minus being attributed to the internal graduation.}
\label{protractor}
\end{center}
\end{figure}

We start by revisiting the notion of  \emph{state}: when it is preceded by the adjective \emph{pure} it is commonly understood within \emph{quantum} formalism as a  unit ray  in  Hilbert space. 
Precisely, a unit ray in the Euclidean plane is an orientation associated to an angle $\phi \in [0, \pi)$, equivalently to a  point of the upper unit semi-circle deleted of  its left boundary point. It is the unit circle $\mathbb{S}^1$ where diametrically opposed points are made identical, i.e. 
\begin{equation}
\label{semcirc}
\mbox{Set of unit rays} \cong \mathbb{S}^1/\{-1,1\}\, . 
\end{equation}
 As a physical realization of this set, we immediately think of the compass without considering the sense, i.e, by identifying North with South, East with West, ...., or even better, to the familiar protractor shown in Figure \ref{protractor}.

The true compass orientations $+$ senses are described by the unit circle itself. In order to avoid notational complications in the present paper, we consider  states as elements of the unit circle, keeping  in mind  the possibility to view as equivalent  two diametrically opposed points.

Now, multiplying a unit vector or a ray by a real number has a deep signification both in Euclidean geometry as in quantum formalism, since it lies at the heart of the superposition principle. The latter states that any linear combination of two vectors is a vector, and that a new unit vector or ray is built from this vector, if it is not zero, by dividing it by its length.  

Next, we revisit the way a state is transformed through the action of a \emph{quantum observable}, i.e, a $2\times2$ symmetric real matrix,   and the way it evolves with the introduction in the formalism of an evolution parameter, i.e. a \emph{time}. We show how such matrices correspond to functions on the circle through a procedure named \textit{integral quantization}. Finally, we show how entanglement and measurement can be understood in this elementary framework.  

A part of the material presented here is borrowed from references \cite{gazeaubook09}, \cite{aagbook13} and \cite{gahell15}. 
The article is organized as follows. In Section \ref{R2dirac} we briefly recall the definition of Hilbert space before rewriting the language of Euclidean geometry by adopting \textit{bra} and \textit{ket} Dirac notations. We then describe \textit{pure} or \textit{mixed}  states, and symmetric $2\times2$ matrices as observables in these real Hilbert spaces. We introduce quantum ``orientation'' observable with two mutually exclusive measurements outcomes, namely ``parallel $=+1$'' and ``perpendicular $=-1$.
In Section \ref{intquant} we proceed with integral quantization of the circle, after a rapid survey of the method. This method allows us to give in a straightforward way the quantum counterparts of classical models such as functions (and even distributions) defined on the circle. These counterparts are $2\times 2$ symmetric real matrices acting on the Euclidean plane.  Some  probabilistic aspects of the obtained quantum objects and of their \textit{semi-classical} portraits are examined in great details. 
Section \ref{evolution} is devoted to the description of a quantum dynamics adapted to our model. Indeed, an adaptation is necessary since we are supposed to use real numbers, vectors and matrices in our quantum formalism, and the latter cannot be derived from the quantization of some  phase space. Such a classical object does not exist here, and so our quantum dynamics cannot be derived from an Hamiltonian system. 
If one wants to use quantum dynamics \textit{\`a la} Schr\"{o}dinger-Heisenberg, one is forced to introduce at a certain stage pure imaginary numbers. Hence, the most general Hamiltonian compatible with the equation ruling the quantum evolution of symmetric $2\times2$  real matrices must be a symmetric complex matrix. 
Nevertheless, we show how to persevere with using real numbers only. This  leads to our Majorana-like  evolution equations.  Then  unitary evolution operators are just  $2\times2$ rotation matrices, as expected in two-dimensional Euclidean geometry. We complete the section with a study of the dynamics of open systems through  the  Lindblad equation. Although non trivial in its full generality, this equation is easily solvable due to the simplicity of our formalism. 
In Section \ref{entanglement}, we address the question of entanglement of two systems with states in the Euclidean plane. States of such objects are  elements of the tensor product $\R^2 \otimes \R^2$. We examine  Bell states and inequality in this context, replacing the usual example of two one-half spins with two orientation systems.    Now, because of the elementary identity $2\times 2= 2+2$, this tensor product is isomorphic to $\R^4= \R^2\times \R^2$ (and also to $\R^2 \oplus \R^2$). Since the plane $\R^2$ can be also viewed as the complex plane  $\C$, we get the possibility of understanding initial entangled objects either as elements of the more elementary complex  Hilbert spaces $\C^2$, e.g., the space of quantum one-half spins, or as elements of the quaternion field $\H$. This set of possibilities sheds an interesting light on entangled objects viewed as cat states in $\C^2$. 
The purpose of Section \ref{magnint} is to give a physical exemple of the quantum Hamiltonian appearing in Euclidean plane quantum dynamics. This is done by considering for instance a  charged spin one half particle, e.g., an electron, interacting with a magnetic field with constant direction in the plane. The demonstration rests upon an integral quantization of the 2-sphere $\mathbb{S}^2$ viewed as the phase space with canonical coordinates $(J_3= J\cos\theta, \phi)$, where $\vec{\mathbf{ J}}$ has the meaning of an angular momentum and $\phi$ is the azimutal angle.  Section \ref{quantmeas} is a description of an elementary quantum measurement of orientations in the plane  in the context of the tensor product of two Euclidean planes, one for the pointer states, the other one for the states of the system.  In Section \ref{conclu} we mention some interesting perpectives of generalisations of the content of our work, centred on the continuation of the game $\R^2 \otimes \R^2 \cong \C^2 \cong \H$ (as vector spaces)  with $\C^2 \otimes \C^2 \cong \C^4 \cong \H \times \H \cong \mathbb{O}$ (as vector spaces), the latter Cartesian product being given an octonionic structure. In Appendix \ref{pararho} we describe different parametrizations of real $2\times 2$ density matrices together with interesting properties. Quaternions in different languages and their basic properties are described  in Appendix \ref{allSU2}, by insisting on  their $\R_+\times\mathrm{SU}(2)$ representation.  In Appendix \ref{unitsphere} we present the integral quantization of the $2$-sphere 
which is needed to establish our result in Section \ref{magnint}.

\section{A  quantum vision of the Euclidean plane}
\label{R2dirac}

\subsection{Euclidean plane with Dirac notations}
\label{dirnotR2}

We first recall  that a  finite-dimensional Hilbert space  is a finite-dimensional real or complex vector space $V$ equipped  with a  scalar (or dot or inner)  product $\lg v| w\rg \in \R$ or $\C$, between any pair of vectors $v$, $w$,   with the following properties 
\begin{align}
\label{hisp1}
  \lg v| w\rg  & = \overline{\lg w| v\rg}   \quad \mbox{complex conjugate symmetry}\, , \\
 \label{hisp2}  \lg v| a_1w_1 + a_2w_2\rg &= a_1\lg v| w_1\rg + a_2  \lg v| w_2\rg   \quad \mbox{linearity in the 2d argument} \, , \\
 \label{hisp3}  \lg v| v\rg &\geq 0\, ,   \quad \mbox{and} \quad \lg v| v\rg = 0 \Rightarrow v=0\, , 
\end{align}
where $a_1$, $a_2$, in \eqref{hisp2} are arbitrary complex or real numbers, and  \eqref{hisp3} encodes the \textit{semi-definiteness} of the scalar product. In the case of a real vector space, the scalar product is fully symmetric and bilinear.   Hence, it is clear that the two-dimensional real vector space equipped with the usual dot product, i.e. the Euclidean plane $\R^2$, is the simplest non-trivial Hilbert space. 

We now adopt the Dirac notations \cite{dirac39} for vectors in  $\R^2$ after restoring the familiar arrow or hat superscripts and boldface symbols. Given the orthonormal basis  of  $\R^2$ defined by the two familiar  unit vectors $\widehat{\boldsymbol{\imath}}$ and $\widehat{\boldsymbol{\jmath}}$, as is shown in Figure  \ref{R2fig}, any vector $\vec{\mathbf{v}}$ in the plane decomposes as
\begin{equation}
\label{vecdec}
\vec{\mathbf{v}} = v_x \widehat{\boldsymbol{\imath}} + v_y \widehat{\boldsymbol{\jmath}}\, .
\end{equation}
We also use  in the sequel the column vector notation to designate $\vec{\mathbf{v}}$ with implicit mention of orthonormal basis, 
\begin{equation}
\label{colvec}
\vec{\mathbf{v}}= \begin{pmatrix}
     v_x     \\
      v_y  
\end{pmatrix}\, , 
\end{equation}
in such a way that the Dirac \emph{ket} and \emph{bra} notations become transparent,
\begin{equation}
\label{ketbradirac}
\vec{\mathbf{v}} = \left| \vec{\mathbf{v}}\right\rg = \begin{pmatrix}
     v_x     \\
      v_y  
\end{pmatrix}\, , \quad \left\lg \vec{\mathbf{v}} \right| = \begin{pmatrix}
     v_x     &
      v_y  
\end{pmatrix}\, . 
\end{equation}
Hence, elementary matrix calculus allows to formulate scalar product and (orthogonal) projector in \textit{bra ket} (respectively in \textit{ket bra}) notations:
\begin{equation}
\label{braketbra}
 \left\lg \vec{\mathbf{v}}\right. \left| \vec{\mathbf{w}}\right\rg = v_xw_x + v_y w_y\, , \quad    \left| \vec{\mathbf{v}}\right\rg  \left\lg \vec{\mathbf{v}} \right| = \begin{pmatrix}
    v_x^2  & v_xv_y    \\
  v_x v_y    &  v_y^2
\end{pmatrix} \, . 
\end{equation}
Since we view the Euclidean plane as a space of quantum states,  pure states are identified with unit vectors, discarding, as announced in the introduction, the imposed equivalence between opposed senses along the orientation support.  Hence, the unit circle $\mathbb{S}^1$ in the plane is the natural set of parameters which distinguish these states. To each $\phi \in [0,2\pi)$, polar angle of the unit vector $\widehat{\mathbf{u}}_{\phi}$, corresponds the pure state $|\phi\rg := \left| \widehat{\mathbf{u}}_{\phi}\right\rg$. In particular, with these notations and associated quantum terminology, the ``horizontal orientation" or \textit{spin up} state is $\widehat{\boldsymbol{\imath}} = |0\rg$, $\widehat{\boldsymbol{\jmath}}=\left| \dfrac{\pi}{2}  \right\rangle$ is the ``vertical orientation" or \textit{spin down} state. 

Their  quality to form an orthonormal basis is encoded in the relations
\begin{equation}
\langle  0 | 0  \rangle = 1 =\left\langle \frac{\pi}{2} \right| \left.\frac{\pi}{2}\right\rangle\, , \quad \langle 0 \left| \frac{\pi}{2} \right\rangle = 0\, ,
\end{equation}
together with the fact that the sum of their corresponding orthogonal projectors \textit{resolves the unity}
\begin{equation}
\label{unity1}
I = | 0  \rangle \langle  0 |  +  \left| \frac{\pi}{2} \right\rangle \left\langle  \frac{\pi}{2} \right| \ \Leftrightarrow \ \left( \begin{array}{cc}
1 & 0\\ 0 & 1 \end{array} \right) = \left( \begin{array}{cc}
1 & 0\\ 0 & 0 \end{array} \right) + \left( \begin{array}{cc}
0 & 0\\ 0 & 1 \end{array} \right)\, .
\end{equation}

\begin{figure}[h!]
\begin{center}
\setlength{\unitlength}{0.15cm} 
\begin{picture}(60,60)
\put(10,20){\vector(1,0){30}} 
\put(10,20){\vector(0,1){30}} 
\put(43, 24){\makebox(0,0){$\widehat{\boldsymbol{\imath}} = |0\rangle \equiv \begin{pmatrix}
      1    \\
      0  
\end{pmatrix}$}}
\put(29, 52){\makebox(0,0){$|\phi\rangle = \begin{pmatrix}
      \cos\phi    \\
      \sin\phi  
\end{pmatrix}$}}  
\put(45, 41){\makebox(0,0){\footnotesize (quantum) state in ``spin representation''}} 
\put(10, 16.5){\makebox(0,0){$O$}} 
\put(30, 10){\makebox(0,0){$\langle  0 | 0  \rangle = 1 =\left\langle \frac{\pi}{2} \right| \left.\frac{\pi}{2}\right\rangle\, , \quad \langle 0 \left| \frac{\pi}{2} \right\rangle = 0$}}
\put(30, 1){\makebox(0,0){$I = | 0  \rangle \langle  0 |  +  \left| \frac{\pi}{2} \right\rangle \left\langle  \frac{\pi}{2} \right| \ \Leftrightarrow \ \left( \begin{array}{cc}
1 & 0\\ 0 & 1 \end{array} \right) = \left( \begin{array}{cc}
1 & 0\\ 0 & 0 \end{array} \right) + \left( \begin{array}{cc}
0 & 0\\ 0 & 1 \end{array} \right)$}} 
\put(8, 52){\makebox(0,0){$\widehat{\boldsymbol{\jmath}} = |\pi/2\rangle =\begin{pmatrix}
      0    \\
      1  
\end{pmatrix}$}} 
\put(16, 38){\makebox(0,0){$1$}} 
\put(20, 26){\makebox(0,0){$\phi$}}
\put(14,20){\oval(10,15)[tr]}
\thicklines 
\put(10,20){\vector(1,2){14}} 
\end{picture}
\end{center}
\caption{The Euclidean plane and its unit vectors viewed as pure quantum states in Dirac ket notations.}
\label{R2fig}
\end{figure}
Pursuing with our quantum terminology,  the  spin representation of a pure state lies in the form of the expansion or decomposition
\begin{equation}
\label{Fock}
| \phi \rangle = \cos{\phi} \,| 0 \rangle  + \sin{\phi}\, \left| \frac{\pi}{2} \right\rangle\, ,
\end{equation}
with its two assumed values
\begin{equation}
\label{Fockval}
\lg 0 | \phi\rg = \cos \phi\, , \qquad \left\lg \frac{\pi}{2} \right|\phi\rg =  \sin\phi\, .
\end{equation}

There exists another representation,  named  ``unit circle'', which is given by the trigonometric function
\begin{equation}
\label{uncirrep}
\lg \eta | \phi\rg = \cos (\phi -\eta)\, .
\end{equation}
However, there is here no quantum   observable underlying localisation on the whole unit circle, at the difference of ordinary quantum mechanics for which there exists a position operator, $Q$, underlying configuration representation, or a momentum operator, $P$,  underlying momentum representation. On the other hand, it is known that in ordinary quantum mechanics  there is no sharp localisation in the corresponding phase space $\{(q,p)\}$ due to the non commutativity of these two operators (Heisenberg!). An analogous situation is encountered here with the fuzzy localisation exemplified by the probability distribution families on the unit circle derived from \eqref{uncirrep},
\begin{equation}
\label{probdistuncirc}
\eta\mapsto \mathcal{P}_{\phi}(\eta) = \vert \lg \eta| \phi\rg\vert^2= \cos^2(\phi-\eta)\, , \quad \int_0^{2\pi} \mathcal{P}_{\phi}(\eta) \, \frac{\ud \eta}{\pi} = 1\, . 
\end{equation}
To the pure state $| \phi\rangle$
 corresponds the orthogonal projector $P_{\phi}$ given by:
\begin{align}
\label{projtheta} 
P_{\phi} &= | \phi \rangle \langle \phi | = \begin{pmatrix}
 \cos{\phi} \\
 \sin{\phi}\end{pmatrix} \begin{pmatrix}
\cos{\phi} &\sin{\phi} \end{pmatrix}  = \begin{pmatrix}
\cos^2{\phi} & \cos{\phi} \sin{\phi}  \\
\cos{\phi} \sin{\phi}  & \sin^2{\phi} \end{pmatrix}\\
\label{projrot} &=\mathcal{R}(\phi) |0\rg\lg 0|\mathcal{R}(-\phi) \, , 
\end{align}
where $\mathcal{R}(\phi)$ is the matrix of rotation in the plane by the angle $\phi$,
\begin{equation}
\label{rotmat}
 \mathcal{R}(\phi): = \begin{pmatrix}
  \cos\phi    &  -  \sin\phi  \\
   \sin\phi   &   \cos\phi 
\end{pmatrix}\,.
\end{equation}
In quantum formalism, the projector $P_{\phi} = | \phi \rangle \langle \phi |$ is also called pure state. This  designation is consistent with the fact that the projector ignores orientation,  $P_{\phi}= P_{\phi + \pi}$.   Then, \textit{mixed states} or \textit{density matrices} are unit trace and  non-negative matrices of the form
\begin{equation}
\label{dens-a-bM}
\rho := \mathsf{M}(a,b) = \begin{pmatrix}
   a   &  b \\
    b  &  1-a
\end{pmatrix}\,,  
\end{equation}
with 
\begin{equation}
\label{a-b}
\quad 0\leq a\leq 1\, , \quad \Delta:= \det \rho = a(1-a)-b^2 \geq 0\, . 
\end{equation}
The conditions \eqref{a-b} express the non-negativeness of $\rho$, i.e.  $\lg \phi | \rho |\phi\rg \geq 0$ for all $|\phi\rg$.   Pure states are rank-one such matrices, i.e., $\rho^2=\rho$,  corresponding to $\Delta = 0$, i.e. $b=\pm \sqrt{a(1-a)}$. More details are given in Subsection \ref{ssecqcirc} and in Appendix \ref{pararho}. A more  enlightening parametrisation is given in terms of the highest eigenvalue $\lambda= \frac{1}{2}(1+\sqrt{1-4\Delta})$, $1/2\leq \lambda \leq 1$  and polar angle $\phi$, $0\leq \phi \leq \pi$ of the corresponding pure eigenstate $|\phi\rg$ pointing toward the upper half-plane.  
In terms of the more suitable parameter $r= 2\lambda -1$, $0\leq r\leq 1$,   the most general form of a real density matrix is given, as a $\pi$-periodic matrix, in terms of the polar coordinates $(r,\phi)$ of a point in the upper half unit disk:
\begin{equation}
\label{standrhomain}
\rho_{r,\phi}= \begin{pmatrix}
  \frac{1}{2}  + \frac{r}{2}\cos2\phi  &   \frac{r}{2}\sin2\phi  \\
\frac{r}{2}\sin2\phi    &   \frac{1}{2}  - \frac{r}{2}\cos2\phi
\end{pmatrix}= \rho_{r,\phi+\pi}\, , \quad 0\leq r\leq 1\, , \ 0\leq \phi < \pi\, . 
\end{equation}
This expression epitomizes the  map from the closed unit disk  to the set of  density matrices. More precisely, each $\rho_{r,\phi}$ is univocally (but not biunivocally) determined by a point in the  unit disk $\mathcal{D}$, with polar coordinates $(r=2\lambda-1, \Phi:=2\phi)$, $0\leq r\leq1$, $0\leq \Phi <2\pi$. 
We check that for  $r=1$  the density matrix is just the orthogonal projector on the unit vector $|\phi\rg$ with polar angle $\phi$, i.e. $ \rho_{1,\phi}= P_{\phi}$.

\begin{figure}
\begin{center}
\includegraphics[width=3in]{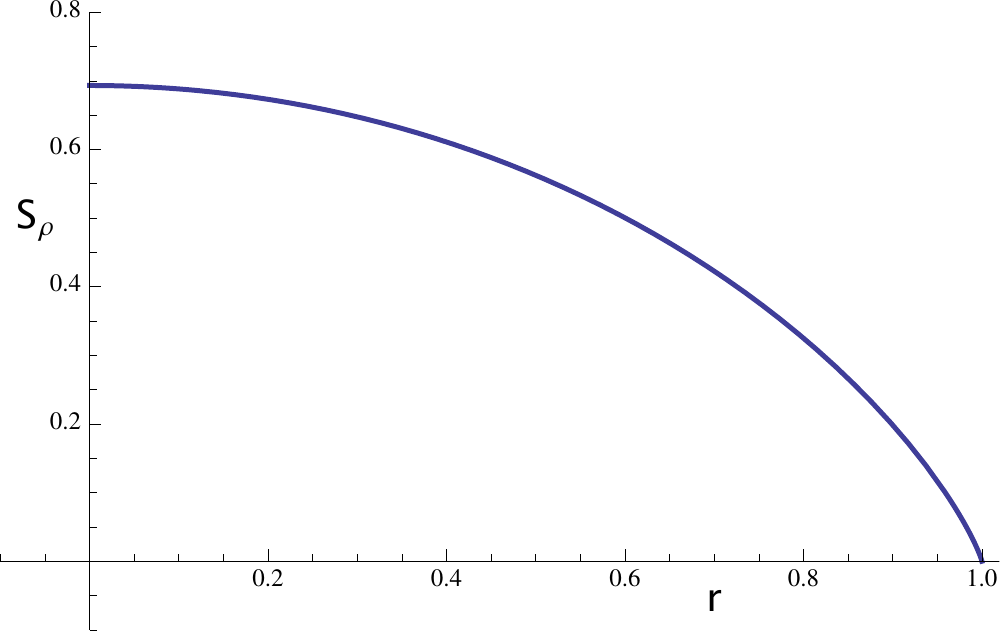}
\caption{Graph of the von Neumann entropy $S_{\rho}$ as a function of parameter $r\in [0,1]$. Its maximal value $\ln 2$ is  reached at $r=0$, which corresponds to the density matrix $\frac{1}{2}I$.}
\label{Srho1}
\end{center}
\end{figure}

The parameter $r$ can be viewed as a measure of the distance of $\rho$ to the pure state $P_{\phi}$ while $1-r$ measures the degree of ``mixing". A statistical interpretation is made possible through the von Neumann entropy \cite{VNentropy,bengtsson07} defined as
\begin{equation}
\label{VNentr}
S_{\rho}:= -\mathrm{Tr}(\rho\,\ln\rho)= -\lambda \ln\lambda - (1-\lambda)\ln(1-\lambda)= -\frac{1+r}{2} \ln\frac{1+r}{2} - \frac{1-r}{2} \ln\frac{1-r}{2}\, .  
\end{equation}
As a function of $\lambda \in [1/2,1]$ $S_{\rho}$ is nonnegative, concave  and symmetric with respect to its maximum value $\log 2$ at $\lambda=1/2$, which corresponds to $r=0$. The graph of $S_{\rho}$ as a function of $r$ is presented in Figure \ref{Srho1}. In  fact, this function is the basic Boltzmann-Gibbs or Shannon entropy in  the case of two possibilities with probabilities $\lambda$ and $1-\lambda$. 

\subsection{Quantum observables}
According to the canons of quantum formalism, a quantum observable is a self-adjoint operator in the Hilbert space of quantum states, because self-adjointness is essential in the Copenhagen-Dirac-von Neumann interpretation of  quantum physics \cite{dirac30}, \cite{vonneumann32}, \cite{omnes94}, \cite{laloe2012} \cite{bricmont16}. Indeed,  the spectral theorem for bounded or unbounded self-adjoint operators is the key for (sharp) quantum measurement, as it was mentioned above.
In our case, a self-adjoint operator $A$ is represented by a $2\times 2$ real symmetric matrix and its spectral decomposition reads in terms of the two orthogonal projectors corresponding to its real eigenvalues: 
\begin{equation}
\label{spmeasR2}
A= \lambda_1 \, P_{\phi_1} + \lambda_2 \, P_{\phi_2}\,, \quad \phi_2= \phi_1 \pm \frac{\pi}{2}\, . 
\end{equation}
An immediate example is provided by the spin representation \eqref{Fock}, which is  also understood  through the  superposition of the two basic projectors
\begin{equation}
\label{locunitcirc}
\Lambda = \lambda_0 |0\rg\lg 0| + \lambda_{\frac{\pi}{2}} \left| \frac{\pi}{2} \right\rangle \left\langle  \frac{\pi}{2} \right|\, . 
\end{equation} 
In terms of quantum measurement, this decomposition  means that the observable ``orientation'' assumes two values only: $\lambda_0$, associated to  the projector on the  horizontal or spin up  state $|0\rg$  and the other one, $\lambda_{\frac{\pi}{2}}$,  associated to  the projector on the vertical or spin down  $\left| \frac{\pi}{2} \right\rangle$. Those measurements of orientation with only two issues are sharp in the sense that  the corresponding projectors are orthogonal.
At this point, we understand why no angular sharp localisation  can be established on the unit circle. By sharp localisation we mean the existence of an angle operator $A$ acting as a multiplication operator in ``angle'' representation $\lg\eta|\phi\rg \equiv \phi(\eta)$. Such an operator does not exist because no $2\times 2$ real symmetric matrix $A$ can satisfy the equation $\lg \eta| A|\phi\rg = \eta\lg\eta|\phi\rg = \eta\cos(\phi-\eta)$, for all $0\leq \phi, \eta, < 2\pi$. 
  
Three basic matrices generate the Jordan algebra \cite{jordan} of all real symmetric $2\times2$ matrices. They are the identity matrix and the two real Pauli matrices,
\begin{equation}
\label{Pauli13}
\sigma_1 = \begin{pmatrix}
   0   &  1  \\
    1  &  0
\end{pmatrix}\, ,  \qquad   \sigma_3 = \begin{pmatrix}
   1   & 0   \\
    0  &  -1
\end{pmatrix}\, . 
\end{equation}
Any element  $A$ of the algebra  decomposes as :
\begin{equation} 
A \equiv\begin{pmatrix}
  a    &   b \\
   b   &  d
\end{pmatrix}  = \frac{a+d}{2}I + \frac{a-d}{2} \sigma_3 + b \sigma_1 
= \alpha I + \delta \sigma_3 + \beta \sigma_1\, . \label{msym13}
\end{equation}
The product in this algebra is defined by
\begin{equation}
\mathcal{O''} = A\odot A' = \frac{1}{2} \left( A A' + A' A\right)\, ,
\label{jord13}
\end{equation}
which entails on the level of components $\alpha, \delta, \beta$, the relations : 
\begin{equation}
\alpha'' = \alpha \alpha' + \delta \delta' + \beta \beta', \  
\delta'' = \alpha \delta' + \alpha' \delta,\  \beta'' = \alpha \beta'+ \alpha' \beta\, .
\end{equation}

Let us associate to the unit vector $\widehat{\mathbf{u}}_{\phi}$ the following symmetric matrix,
\begin{equation}
\label{sigphi}
\sigma_{\phi}:=  \cos \phi\, \sigma_{3} +  \sin \phi\, \sigma_1\equiv \overrightarrow{\boldsymbol{\sigma}}\cdot \widehat{\mathbf{u}}_{\phi}= \begin{pmatrix}
  \cos \phi    &  \sin \phi  \\
 \sin \phi     & - \cos \phi
\end{pmatrix} = \mathcal{R}(\phi)\,\sigma_3\, ,
\end{equation}
where we have introduced the unusual vector notation $\overrightarrow{\boldsymbol{\sigma}}= \sigma_3 \widehat{\boldsymbol{\imath}}+ \sigma_1\widehat{\boldsymbol{\jmath}}$.  
Thus, this operator  (see also \eqref{rotmatsig}) is the composition of the rotation by angle $\phi$ with the mirror symmetry with respect to the polar axis. We will use it in Section  \ref{entanglement} for presenting  the Bell inequality within the context of the Euclidean plane. 
Its important feature of \eqref{sigphi} lies in its spectral decomposition, 
\begin{equation}
\label{sig2phi}
\sigma_{\phi} = \left|\frac{\phi}{2}\right\rg\left\lg\frac{\phi}{2}\right| - \left| \frac{\phi}{2} + \frac{\pi}{2}\right\rg\left\lg \frac{\phi}{2} + \frac{\pi}{2}\right|\, .
\end{equation}
It leads us to interpret  $\sigma_{\phi}$ as the quantum observable associated to the orientation ``polar angle" $\phi/2$ (or the equivalence class of lines parallel to $\widehat{\mathbf{u}}_{\phi/2}$),  corresponding to the measurement outcome $+1$ (i.e., parallel), and its orthogonal in the plane (or the set of  lines normal to $\widehat{\mathbf{u}}_{\phi/2}$), corresponding to the outcome $-1$ (i.e., perpendicular). Note the actions of  $\sigma_{\phi}$ on states $|\phi\rg$ and its orthogonal $|\phi +\pi/2\rg$, and   the average value of  $\sigma_{\phi}$ in state $|\theta/2\rg$, 
\begin{equation}
\label{avasig}
\sigma_{\phi}|\phi\rg = |0\rg\,, \quad \sigma_{\phi}\left|\phi + \frac{\pi}{2}\right\rg = -\left|\frac{\pi}{2}\right\rg\, , \quad  \left\lg\frac{\theta}{2}\right| \sigma_{\phi} \left|\frac{\theta}{2}\right\rg = \cos (\theta-\phi) = \lg \theta|\phi\rg=\widehat{\mathbf{u}}_{\theta}\cdot\widehat{\mathbf{u}}_{\phi}\, . 
\end{equation}

It is interesting to notice that the density operator \eqref{standrhomain} can be expressed in terms of $\sigma_{\phi}$ as
\begin{equation}
\label{standrhomain1}
\rho_{r,\phi}= \frac{1}{2}\left( I + r \sigma_{2\phi}\right)\, . 
\end{equation}

\section{Integral quantization of the circle}
\label{intquant}

It is mostly accepted that a quantum model for a system has a classical counterpart from which it is built through some quantization procedure. In this section, we present a straightforward quantization based on integral calculus. Then we explain how the classical counterparts of the above $2\times2$ symmetric matrices, viewed as quantum observables, are the quantized versions of real-valued functions on the circle, and that there is a one-to-one  relation if we restrict the functions to be linear in  elements in the finite set $\{1, \cos{2\phi}, \sin{2\phi}\}$. 

\subsection{An outline of integral quantization}

Any quantization of functions defined on a set $X$ (e.g. a phase space
or something else)  should reasonably meet four basic requirements \cite{aagbook13}:%
\subsubsection*{(i) Linearity}
It is a linear map $f\mapsto A_f$:
\begin{equation} \label{qmap1}
\mathfrak{Q}:\mathcal{C}(X)\mapsto\mathcal{A}(\calH)\, , \qquad \mathfrak{Q}(f) = A_{f}\, , 
\end{equation} 
where
\begin{itemize}
  \item $\mathcal{C}(X)$ is a vector space of complex or real-valued functions $f(x)$
on a set $X$, i.e. a ``classical'' mathematical model, 
\item $\mathcal{A}(\calH)$ is a vector space of linear operators  in some real or complex Hilbert space $\calH$, i.e, a ``quantum'' mathematical model. 
\end{itemize}
The map \eqref{qmap1} is 
 such that 
\subsubsection*{(ii) Unity}
the function $f=1$ is mapped to the identity operator $I$ on $\calH$, 
\subsubsection*{(iii) Reality}
a real $f$ is mapped to a  self-adjoint operator $A_{f}$
in $\calH$ or, at least, a symmetric operator (in the infinite dimensional case) 
\subsubsection*{(iv) Covariance}
if $X$ is acted on by a symmetry group $G$, i.e. for $g\in G$, $G\ni x\mapsto g\cdot x\in X$, then there exists a unitary representation $U$ of the group such that  $A_{T(g)f}= U(g)A_f U(g^{-1})$, with $(T(g)f)(x)= f\left(g^{-1}\cdot x\right)$.
 
In our case, $\calH$ is finite-dimensional and the operators $A_f$ are viewed as matrices. 
Of course, further requirements are necessarily added, depending on
the mathematical structures equipping $X$ and $\mathcal{C}(X)$  and their respective physical relevance.

An easy way to achieve the above minimal programme is to use resources offered by some adapted integral calculus. Suppose that the set $X$ is equipped with a measure $\ud\nu(x)$ allowing to deal with integrals like 
$\int_{X}T(x)\,\ud\nu(x)$, in which the functions $T(x)$ are matrix valued, the integral being performed on each matrix element.  Now, suppose that there exists an $X$-labeled
family of matrices $M(x)$ resolving the
identity $I$ in the following sense: 
\begin{equation}
\label{resunitM}
X\ni x\mapsto\mathsf{M}(x)\,,\quad\int_{X}\,{\sf M}(x)\,\ud\nu(x)=I\,
\end{equation} 
The case we are specially interested in what occurs when the ${\sf M}(x)$'s are unit trace and non-negative, i.e. are density matrices, 
\begin{equation} \label{densop}
{\sf M}(x)=\rho(x)\,, 
\end{equation} 
because of their possible probabilistic content (see below). 
With this material at hand, the integral quantization of  functions $f(x)$ on $X$ is the
linear map 
\begin{equation}
\label{iqmap}
f\mapsto A_{f}=\int_{X}\,{\sf M}(x)\, f(x)\,\ud\nu(x)\,,
\end{equation}
which fulfills at least our quantization requirements (i) and (ii) and it takes the classical function $f(x)$ to the operator $A_f$. If the matrices ${\sf M}(x)$ are real symmetric, then (iii) is also satisfied. To get (iv), i.e. covariance, we need to add more structure on the measure space $(X,\nu)$. Such structures appear in a natural way in the examples that we consider in this paper, namely the circle, $X= \mathbb{S}^1$, which is considered in the sequel,  and the sphere, $X= \mathbb{S}^2$, which is considered in Appendix \ref{unitsphere}.

\subsection{Semi-classical portraits}

Any quantization map should be complemented with a ``dequantization" map, associating with $A_f$ a function $\check f(x)$, more or less similar to the original $f(x)$, and giving a ``semi-classical portrait" of $A_f$. In quantum mechanics, examples are provided by the Husimi function or the Wigner function\cite{zachos05}, more generally  by  lower (Lieb)  \cite{lieb73} or contravariant (Berezin) \cite{berezin75} symbols. The terminology \textit{lower}, which will appear below together with \textit{upper}, is justified by the two Berezin-Lieb inequalities which follow from the symbol formalism. 

As a matter of fact, some  properties of the operator $A_{f}$, may be derived or at least well grasped from functional properties
of its lower symbol, which is defined in the present context by: 
\begin{equation}
 \label{lowsymbgen}
A_{f}\mapsto\check{f}(x):=\mathrm{tr}(\mathsf{M}(x)\, A_{f})= \int_{X}\,\mathrm{tr}(\mathsf{M}(x){\sf M}(x^{\prime}))\, f(x^{\prime})\,\ud\nu(x^{\prime})\,.
\end{equation} 
When $\mathsf{M}(x)=\rho(x)$ (density matrix), the map $x^{\prime} \mapsto \mathrm{tr}(\rho(x)\rho(x^{\prime}))$ defines a probability
distribution $\mathrm{tr}(\rho(x)\rho(x^{\prime}))$ on the measure space $(X,\ud\nu(x^{\prime}))$, as it is easily proved by multiplying the resolution of the unity in \eqref{resunitM} with $\rho(x)$ and tracing the result. Then the function $\check{f}(x)$ is the
local average of the original $f$ with respect to the above distribution:
\begin{equation}
 \label{locaver}
f(x)\mapsto\check{f}(x)=\int_{X}f(x^{\prime})\,\mathrm{tr}(\rho(x)\rho(x^{\prime}))\,\mathrm{d}\nu(x^{\prime})\,.
\end{equation} 
In many cases, this generalisation of convolution has a regularising effect on the original $f$. For instance, if $f(x)= \delta_{x_0}(x)$ is the Dirac function peaked at $x_0\in X$, 
\begin{equation}
\label{diracx0}
\int_{X}g(x)\,\delta_{x_0}(x)\,\mathrm{d}\nu(x)= g(x_0)\,, 
\end{equation} 
for any test function $g(x)$, then $\check{\delta}_{x_0}(x)=\mathrm{tr}(\rho(x)\rho(x_0))$, a function which is expected to be  more regular around $x_0$.

\subsection{Quantization of the circle}
\label{ssecqcirc}
In the context of the present work, the measure space $(X,\ud \nu(x))$ is the unit circle equipped with its uniform (Lebesgue) measure:
\begin{equation}
\label{measS1}
X= \mathbb{S}^1\, , \quad \mathrm{d}\nu(x) = \frac{\ud \phi}{\pi}\, , \quad \phi \in [0, 2\pi)\, .
\end{equation}
 The Hilbert space is the Euclidean plane $\mathcal{H}= \R^2$. Note that $\mathbb{S}^1$ is also a group, isomorphic to the abelian group SO$(2)$ of rotations in the plane. Thus, we can expect that rotational covariance mentioned in point (iv) plays a central role here. It is indeed the case since the covariance property \eqref{rotrho} described in Appendix \ref{pararho} allows one to define the family of density operators obtained from rotational transport of the initial density operator introduced in \eqref{standrhomain},  
\begin{equation}
\label{rotrhomain}
 \rho_{r,\phi_0}(\phi)= \mathcal{R}\left(\phi\right) \rho_{r,\phi_0}\mathcal{R}\left(-\phi\right)=  \rho_{r,\phi_0+\phi}\, , \quad 0\leq \phi< 2\pi\, . 
\end{equation}
where the rotation matrix $\mathcal{R}(\phi) $ is defined by \eqref{rotmat}.  It is straightforward to prove that this family resolves the identity
\begin{equation}
\label{margomegamain}
\int_0^{2\pi} \rho_{r,\phi_0}(\phi) \, \frac{\ud\phi}{\pi}= I\,.
\end{equation}
The proof can be made by hand, integrating each matrix element. Actually it is the direct consequence of a famous result in group representation theory, named Schur's Lemma \cite{barracz77}. 
It follows the $\mathbb{S}^1$-labelled family of probability distributions on $(\mathbb{S}^1\,, \, \ud \phi/\pi)$
\begin{equation}
\label{probdistcirc}
 p_{\eta}(\phi) = \mathrm{tr}\left(\rho_{r,\phi_0}(\eta)\,\rho_{r,\phi_0}(\phi)\right)= \frac{1}{2}\left(1 + r^2\cos 2(\phi-\eta)\right)\, . 
\end{equation}
Such an expression  reminds us of the cardioid distribution (see \cite{mardia72}, page 51). 
At $r= 0$ we get the uniform probability on the circle whereas at $r=1$ we get the simplified resolution of the identity
\begin{equation}
\label{resuntheta}
\int_0^{2\pi} |\phi\rg\lg \phi| \, \frac{\ud\phi}{\pi}= I\,.
\end{equation}
and the pure state probability distribution  $\mathcal{P}_{\eta}(\phi) =\cos^2\left(\phi-\eta\right)$ previously introduced in \eqref{probdistuncirc}. 

The quantization of a function (or distribution) $f(\phi)$ on the circle based on \eqref{margomegamain}   leads to 
the 2$\times$2 matrix operator
\begin{equation}
\label{qtfrhor}
f \mapsto A_f = \int_0^{2\pi} f(\phi) \rho_{r,\phi_0}(\phi) \, \frac{\ud\phi}{\pi}= \begin{pmatrix}
  \lg f\rg  + \frac{r}{2}C_c\left(R_{-\phi_0}f\right)  &   \frac{r}{2}C_s\left(R_{-\phi_0}f\right) \\
\frac{r}{2}C_s\left(R_{-\phi_0}f\right)   &   \lg f\rg - \frac{r}{2}C_c\left(R_{-\phi_0}f\right)
\end{pmatrix}\,,
\end{equation}
where $ \lg f\rg:= \frac{1}{2\pi}\int_0^{2\pi}f(\phi)\,\ud\phi$ is the average of $f$ on the unit circle and $R_{\phi_0}(f)(\phi) := f(\phi-\phi_0)$. The symbols $C_c$ and $C_s$ are for the cosine and sine doubled angle Fourier coefficients of $f$,
\begin{equation}
\label{CcCs}
C_c(f) = \int_0^{2\pi} f(\phi) \cos2\phi \, \frac{\ud\phi}{\pi}\, , \quad C_s(f) = \int_0^{2\pi} f(\phi) \sin2\phi \, \frac{\ud\phi}{\pi}\, . 
\end{equation}
The eigenvalues of the symmetric matrix $A_f$ are 
\begin{equation}
\label{eigenAf}
\lambda_{\pm} =  \lg f\rg \pm  \frac{r}{2} \sqrt{C^2_c\left(R_{-\phi_0}f\right) + C^2_s\left(R_{-\phi_0}f\right)}\, , 
\end{equation}
with  respective orthogonal eigenvectors
\begin{equation}
\label{eigenvAf}
|\phi_{\pm}\rg\, , \quad \tan \phi_{\pm} = \frac{C_s\left(R_{-\phi_0}f\right)}{C_c\left(R_{-\phi_0}f\right) + \lambda_{\pm}}= \frac{C_c\left(R_{-\phi_0}f\right) - \lambda_{\pm}}{C_s\left(R_{-\phi_0}f\right) }\, . 
\end{equation}
The covariance property, i.e., our 4th requirement on any quantization procedure, is fulfilled here by application of \eqref{rotrhomain}:
\begin{equation}
\label{covpropcircle}
 \mathcal{R}\left(\theta\right) \,A_f\,\mathcal{R}\left(-\theta\right)= A_{R_{\theta}(f)}\,. 
\end{equation}
  
The simplest function to be quantized is the angle function $\Da(\phi)$, i.e. the  $2\pi$-periodic extension of $\Da(\phi)  = \phi$, up to the addition of a constant,  for $\phi \in [0,2\pi)$,
\begin{equation}
\label{qtfrhora}
 A_{\Da}= \begin{pmatrix}
  \pi  +  \frac{r}{2}\sin 2\phi_0  &   - \frac{r}{2}\cos2\phi_0 \\
- \frac{r}{2}\cos2\phi_0   &  \pi -  \frac{r}{2}\sin2\phi_0
\end{pmatrix}= \pi I + \frac{r}{2}\sigma_{2\phi_0 - \pi/2} \,,
\end{equation}
with the notations of \eqref{sigphi}. 
 Its eigenvalues are $\pi   \pm \dfrac{r}{2}$ with corresponding  eigenvectors $\left|\phi_0 \mp\dfrac{\pi}{4}\right\rg$. 
Its lower symbol is given by the smooth function
\begin{equation}
\label{lowsagluc}
\check{\Da}(\phi) = \pi - \frac{r^2}{2}\sin2\phi\,, 
\end{equation}
which reduces to the constant $\pi$ for $r=0$, i.e., to the average of the angle function for the uniform distribution on the circle. 

It is natural to ask what is a  classical counterpart of the  quantum orientation observable $\sigma_{\theta}$ (for the direction state $|\theta/2\rg$) introduced in \eqref{sigphi} and \eqref{sig2phi}. It is the quantization through the map \eqref{qtfrhor} of the following function 
\begin{equation}
\label{qsigphi}
f(\phi)=\frac{2}{r} \widehat{\mathbf{u}}_{\theta+\phi_0}\cdot\widehat{\mathbf{u}}_{\phi}\mapsto A_f= \sigma_{\theta}\, . 
\end{equation}

\subsection{Quantization with pure states and symbol calculus}

The existence of the set  $\{ | \phi \rangle\lg \phi| \}$ as the particular case $r=1$ and $\phi_0=0$ of the density matrices \eqref{rotrhomain} permits to proceed with the simplified ``coherent state"  integral quantization  
\begin{equation}
f \mapsto A_f = \frac{1}{\pi} \int_0^{2\pi} \ud\phi \, f(\phi) | \phi \rangle
\langle \phi |\, . 
\label{quantcircle13}
\end{equation}
  The quantized version of the angle function  $\Da (\phi)$ introduced above becomes:
\begin{equation}
\label{qangle13}
A_{\Da } = \begin{pmatrix}
  \pi    &  -\frac{1}{2}  \\
    -\frac{1}{2}    &  \pi
\end{pmatrix}\, ,
\end{equation}
with eigenvalues $\pi \pm \frac{1}{2} $. Thus, a localisation measurement on the circle would yield those two values only, along the spectral decomposition
\begin{equation}
\label{spdecangle}
A_{\Da } = \left(\pi + \frac{1}{2}\right)\left\vert  \frac{3\pi}{4}\right\rg\left\lg \frac{3\pi}{4}\right\vert + \left(\pi - \frac{1}{2}\right)\left\vert\frac{\pi}{4}\right\rg\left\lg  \frac{\pi}{4}\right\vert\, ,  
\end{equation}
with 
\begin{equation}
\label{vpangle}
\left\vert\frac{\pi}{4}\right\rg\left\lg  \frac{\pi}{4}\right\vert= I+\sigma_1\, , \quad \left\vert\frac{3\pi}{4}\right\rg\left\lg  \frac{3\pi}{4}\right\vert= I-\sigma_1\, . 
\end{equation}
The family of pure states  $\{ | \phi \rangle\lg \phi|  \}$  allows one to carry out the symbol calculus \emph{\`a la} Berezin-Lieb mentioned above. To any \emph{self-adjoint} operator $A$ one  associates the two types of \emph{symbol}, functions $\check{A}(\phi)$ et $\widehat{A}(\phi)$ respectively defined on the unit circle by 
\begin{equation}
\label{lowsymb13}
\check{A}(\phi) = \langle  \phi  | A| \phi  \rangle \,  : \
\mbox{lower or covariant symbol}\, ,
\end{equation} 
 and 
\begin{equation}
A = \frac{1}{\pi} \int_0^{2\pi} \ud\phi \, \widehat{A}(\phi) | \phi  \rangle
\langle \phi |\, . \label{ssup13}
\end{equation}
where $\widehat{A}(\phi)$ is named upper or contravariant symbol of $A$, besides the fact that it is also the classical counterpart of $A$ in the context of integral quantization.    
 
The simplest upper symbols  and the lower symbols of non-trivial basic elements are respectively given by : 
\begin{equation}
\label{symsigma}
\cos{2\phi} =  \check{\sigma}_3 (\phi) = \frac{1}{2}\widehat{\sigma}_3 (\phi),\ \sin{2\phi} = 
\check{\sigma}_1 (\phi) = \frac{1}{2}\widehat{\sigma}_1 (\phi)\, .
\end{equation}
For the symmetric matrix  (\ref{msym13}) there follows the two symbols :
\begin{align}
\label{symbop13A}
\check{A}(\phi) &= \frac{a+d}{2} + \frac{a-d}{2} \cos{2\phi} + 
b \sin{2\phi} = \alpha + \delta \cos{2\phi} + \beta \sin{2\phi}\, , \\
\label{symbop13B} \widehat{A}(\phi) & =\alpha + 2\delta \cos{2\phi} + 2\beta \sin{2\phi} = 2 \check{A}(\phi) - \frac{1}{2}\mathrm{tr}\,A\, .
\end{align}

\begin{figure}[!htb]
\centering
\includegraphics[width=4in]{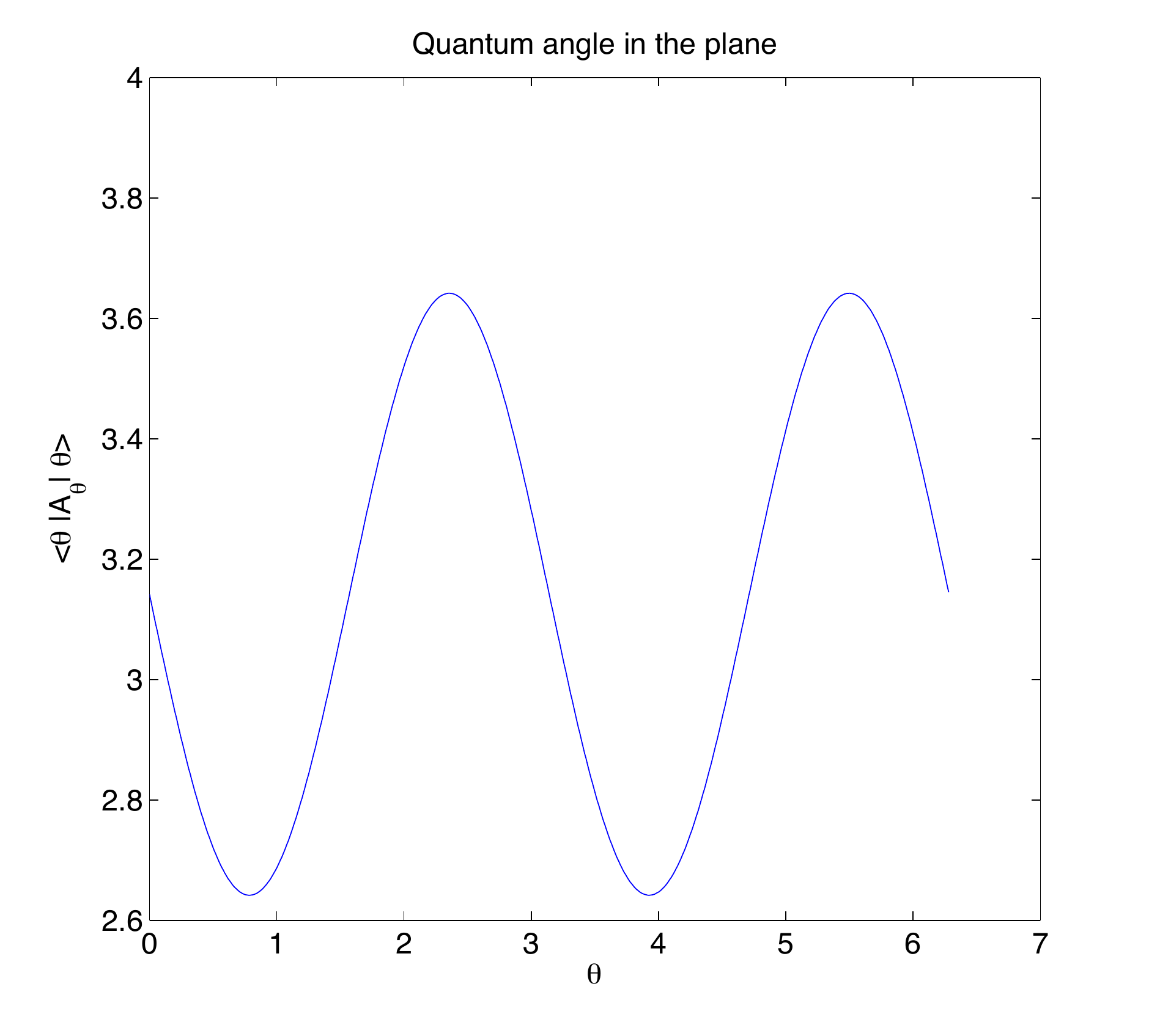}
\caption{The lower symbol  of the angle operator in the two-dimensional quantization of the circle is the $\pi$-periodic function $\pi  - \frac{1}{2} \sin{2\phi}$ that  varies between the two eigenvalues of $A_{\phi}$.}
\label{quantangle13}
\end{figure}

For instance, the lower symbol of the quantum angle (\ref{qangle13}) is obtained by fixing $r=1$ in \eqref{lowsagluc}:
\begin{equation}
\label{symbangle13}
\langle \phi | A_{\Da }|\phi \rangle = \pi  - \frac{1}{2} \sin{2\phi},
\end{equation}
a $\pi$-periodic function that smoothly varies between the two eigenvalues of $A_{\phi}$, as is shown in Figure \ref{quantangle13}.

As announced in the preamble of this section, one notices that  all the  symbols defined above belong to  the subspace $\mathcal{V}_{A}$ of real Fourier series that is  the linear span of the three functions $1$, $\cos{2\phi}$, $\sin{2\phi}$. Hence $\widehat{A}(\phi)$ is highly non unique since it is defined up to the addition of any function $N(\phi)$ lying in the orthogonal complement of $\mathcal{V}_{A}$, i.e.  such that $ \int_0^{2\pi} \ud\phi \, N(\phi) | \phi  \rangle\langle \phi |= 0$.

For example,  the expressions of the upper and lower symbols of the density matrix \eqref{standrhomain1} read as 
\begin{equation}
\label{standrhomainlus}
\widehat{\rho}_{r,\phi}(\theta)= \frac{1}{2}+  r \cos 2(\theta- \phi)\, , \quad \check{\rho}_{r,\phi}(\theta)= \frac{1}{2}(1+  r \cos 2(\theta- \phi))  \, . 
\end{equation}

Let us finally make explicit the Berezin-Lieb inequalities in the present context.  Let $g$ be a convex function. Denoting by  $\lambda_{\pm}$ the eigenvalues of the symmetric matrix
 $A$, we have
\begin{equation}
\frac{1}{\pi} \int_0^{2\pi} g(\check{A}(\phi)) \, \ud\phi  \leq \mathrm{tr}(g(A)) 
= g(\lambda_+) + g(\lambda_-) \leq \frac{1}{\pi} \int_0^{2\pi} g(\widehat{A}(\phi)) \, \ud\phi\, .
\end{equation}
This double inequality justifies the use of adjectives \textit{lower} and \textit{upper} for symbols.

\subsection{Probabilistic aspects}
\label{probasp13} 
Behind the resolution of the identity  \eqref{resuntheta} lies an interesting interpretation in terms of geometrical probability.  Let us consider a  Borel \cite{borelset} subset $\Delta$ of the interval $\lbrack 0, 2\pi)$ and the  restriction to $\Delta$ of the  integral \eqref{resuntheta}:
\begin{equation}
a(\Delta) = \frac{1}{\pi} \int_{\Delta} \ud\phi \, | \phi  \rangle \langle
 \phi  |\, . \label{povp13}
\end{equation}
One easily verifies the following properties :
\begin{align}
\label{propgeom13}
\nonumber & a(\emptyset) = 0\, , \quad  a(\lbrack 0, 2\pi)) = I_d, \\
&  a(\cup_{i\in J}\Delta_i) = \sum _{i\in J}a(\Delta_i)\, , \quad \mbox{if} \ \Delta_i \cap \Delta_j = \emptyset \  \mbox{for
all}\  i\not = j\,  .
\end{align}
The application $\Delta \mapsto a(\Delta)$ defines  a normalized measure on the  $\sigma$-algebra \cite{sigmaalg} of the Borel sets in the interval $\lbrack 0, 2\pi)$, assuming its values in the set of  positive or null $2\times 2$ matrices \index{POVM} (POVM). 

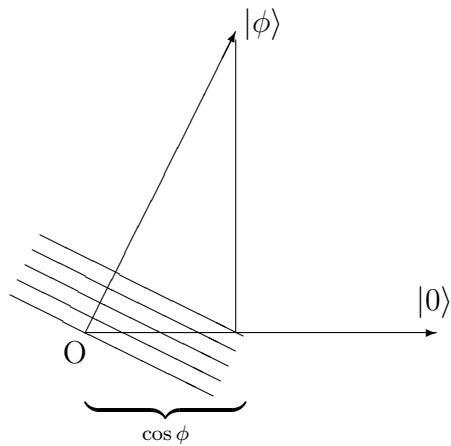
\begin{figure}[h]
\begin{center}
\unitlength=1cm
\begin{picture}(8,6)
\put(5.1,4.9){\line(0,-1){3.9}}
\put(2.8,.6){O}
\put(7.5,1.3){$ | 0  \rangle$}
\put(5.2,5){$ | \phi  \rangle$}
\put(3.1,1){\vector(1,2){2}}
\put(3.1,1){\vector(1,0){4.67}}
\multiput(2.1,1.5)(.1,.2){5}{\line(2,-1){2.7}}
\put(3.1,.1){$\underbrace{\ \ \ \ \ \ \ \ \  \ \ \ \ \ \ \, }_{\cos{\phi}}$}
\end{picture}
\end{center}
\vskip 0.5cm
\caption{Set $\{ D_{\phi, p} \}$ of straight lines normal to $| \phi  \rangle$  
that intersect the  segment  with origin $O$ and length $\vert \cos{\phi} \, \vert$ equal to the  projection of $|
\phi  \rangle$  onto $ | 0  \rangle$.}
\label{probge13}
\end{figure}

Let us now put into evidence the probabilistic nature of the measure $a(\Delta)$.  Let $| \eta \rangle$ be a unit vector. The application 
\begin{equation}
\Delta \mapsto \langle  \eta | a(\Delta) | \eta \rangle = \frac{1}{\pi} \int_{\Delta} \cos^2(\phi - \eta) \, \ud\phi
\end{equation}
is clearly a probability  measure, since it involves precisely the  distribution $\mathcal{P}_{\phi}(\eta)$ introduced in  \eqref{probdistuncirc}. For instance, for $\Delta = [\theta_1,\theta_2]$ and $\eta = 0$, we have
\begin{equation}
\label{exaDelta}
a\left([\theta_1,\theta_2]\right)= \frac{\theta_2-\theta_1}{2\pi} - \frac{1}{\pi}\sin\left(\theta_2-\theta_1\right) \sin\left(\theta_2+\theta_1\right)\, . 
\end{equation} 
Now,  the quantity $\langle  \eta  | a(\Delta) | \eta \rangle$
means that  direction $| \eta  \rangle$ is examined from the point of view of the family of vectors $\left\{| \phi \rangle, \,   \,  \phi\in \Delta\right\}$. As a matter of fact, it has a  \emph{geometrical probability} interpretation in the  plane \cite{delt}. With no loss of generality let us choose  $\eta = 0$.
Recall here the canonical equation describing a straight line   $D_{\phi, p}$ in the plane  :
\begin{equation}
\langle  \phi   |\overrightarrow{\mathbf{v}}  \rangle = \cos{\phi}\, x + \sin{\phi}\, y = p\, ,\quad \overrightarrow{\mathbf{v}} = \begin{pmatrix}
      x    \\
      y  
\end{pmatrix}\, , 
\end{equation}
where $| \phi\, \rangle$ is the direction normal  to $D_{\phi, p}$ and the parameter $p$ is equal to the distance of $D_{\phi, p}$ to the origin.
 There follows that $\ud p\,\ud\phi$ is the (non-normalized) probability measure element on the set $\left\lbrace D_{\phi, p} \right\rbrace$ of the lines randomly chosen in the plane. 
Picking a certain $\phi$, consider the set $\{ D_{\phi, p} \}$ of the lines normal to $| \phi  \rangle$ that intersect
the segment  with origin $O$ and length $\vert \cos{\phi} \vert$ equal to the projection
of $| \phi  \rangle$  onto  $ | 0  \rangle$ as shown in Figure \ref{probge13} 

The measure of this set is equal to :
\begin{equation} 
\label{measline13}
 \left(\int_{0}^{\cos^2\phi} \ud p\right) \, d\phi = \cos^2\phi \,\ud\phi \, .
\end{equation}
Integrating (\ref{measline13}) over all directions  $| \phi  \rangle$ gives the area of the unit circle. Hence we can construct 
$\langle  \eta  | a(\Delta) | \eta  \rangle$ as the probability 
for a straight line in the plane to belong to the set of secants of segments that are projections  
$\langle  \eta  | \phi \rangle$ of the unit vectors  $| \phi  \rangle$, $\phi \in \Delta$, onto the unit  vector 
 $| \eta  \rangle$. 
 
%

\section{Time evolution, rotation, and Baby Majorana equations}
\label{evolution}
\subsection{Closed system}
We now deal with the question of the existence of a quantum dynamics  compatible with the simplicity of our model. Let us introduce into the game a time (or evolution) parameter $t$, which is of classical nature. Following the principles established by Heisenberg and Schr\"{o}dinger \cite{dirac30},  the quantum dynamics of a system rests upon the existence of the associated, possibly time-dependent,  Hamiltonian operator $H= H(t)$. Given some initial time $t_0$, this operator generates its corresponding  evolution operator  $U(t,t_0)$ through the equation
\begin{equation}
\label{eveq}
\ii \hbar \frac{\partial}{\partial t} U(t,t_0)= H(t)\, U(t,t_0)\, .
\end{equation}
The operator $U$ is unitary,
\begin{equation}
\label{unitary} U(t,t_0)U(t,t_0)^{\dag} = U(t,t_0)^{\dag}U(t,t_0)= I\, . 
\end{equation}
Note that  these two above relations imply that $H(t)\, U(t_0,t)= U(t_0,t)\,H(t)$. 
The operator $U$ is expected to obey the composition rule
\begin{equation}
\label{compos} U(t_2,t_1)U(t_1,t_0) = U(t_2,t_0)\, , \quad t_2>t_1>t_0\,, 
\end{equation}
which implies $U(t,t)= I$ for all $t$. 
As representing a physical quantity, namely the energy of the system,  the Hamiltonian is a self-adjoint operator. From \eqref{eveq} are derived two celebrated equations, the Schr\"{o}dinger one when this relation between operators is applied to a quantum state, 
\begin{equation}
\label{schroding}
\ii \hbar \frac{\partial}{\partial t} U(t,t_0)|\psi\rg\equiv \ii \hbar \frac{\partial}{\partial t}|\psi(t)\rg=  H(t)\, |\psi(t)\rg\, ,\quad |\psi\rg= |\psi(t_0)\rg\, , 
\end{equation}
and the Heisenberg-Dirac equation when applied to a possibly time-dependent quantum observable $A(t)$, 
\begin{equation}
\label{heisenb}
\ii \hbar \frac{\ud}{\ud t} A_H= [A_H, H] +  U(t,t_0)^{\dag}\left(\ii \hbar \,\frac{\partial}{\partial t}A\right)U(t,t_0) \, , 
\end{equation}
where $A_H(t):= U(t,t_0)^{\dag} A(t)U(t,t_0)$. 

In the  present context of the Euclidean plane, can we involve a $2\times2$ real symmetric matrix as a Hamiltonian $H$?  A first problem concerns the construction of $H$ without the existence of a classical phase space as a guidance. Let us circumvent  this difficulty by simply postulating its existence. Then, another issue makes our effort totally pointless.  Indeed, from  the equations \eqref{unitary} and \eqref{compos} we infer that the evolution operator must be the rotation matrix
\begin{equation}
\label{unrot}
U(t,t_0) = \mathcal{R}(\phi(t) - \phi(t_0))
\end{equation}
for a certain function $\phi(t)$, and if we rely on \eqref{eveq}, we \underline{must} deal with a $2\times 2$ Hermitian complex
 matrix
 \begin{equation}
\label{Hcom}
H(t)= \begin{pmatrix}
   a(t)   & b(t)    \\
   \overline{b(t)}   &  d(t)
\end{pmatrix}\, , \quad \mbox{with}\quad  a(t)\,,\, d(t)\in \R\,,\, b(t) \in \C\,, 
\end{equation}
in order to have a non-trivial solution. We then  solve \eqref{eveq}  to find for the above coefficients 
\begin{equation}
\label{hamsol}
a(t) = 0 =d(t)\, ,  \quad \ii \hbar \frac{\ud \phi}{\ud t} = b(t)\equiv -\ii \mcE(t)\, ,   
\end{equation}
Hence the allowed form of the  Hamiltonian reads
\begin{equation}
\label{hamsol1}
H(t)= \begin{pmatrix}
   0   & -\ii  \mcE(t)  \\
    \ii  \mcE(t)  &  0
\end{pmatrix} = \mcE(t)\sigma_2\, ,
\end{equation}
and  for the evolution operator
\begin{equation}
\label{evrot}
U(t,t_0) = \mathcal{R}\left(\frac{1}{\hbar}\int_{t_0}^{t}\mcE(t^{\prime})\,\ud t^{\prime}\right)\, . 
\end{equation}
Hence the presence of the  imaginary unit $\ii$ in \eqref{Hcom} is unavoidable. If we want to keep the reality of our Euclidean plane and of the matrices acting on it,  no real symmetric matrix, except the null matrix, can represent an Hamiltonian in the  context of our approach. If we wish to keep the dynamics consistent with   our formalism based on real numbers only, we have to abandon the requirement for $H$ to be symmetric. Instead, let us introduce the  real ``pseudo-Hamiltonian''
\begin{equation}
\label{pseudoH}
\widetilde H(t)= \begin{pmatrix}
   0  & -\mcE(t)   \\
   \mcE(t)   &  0
\end{pmatrix} = \mcE(t)  \tau_2\, , \quad \tau_2= -\ii \sigma_2= \begin{pmatrix}
   0   & -1   \\
   1   &  0
\end{pmatrix}\, .
\end{equation}
Then, by dividing by $\ii$ the evolution equations \eqref{schroding} and \eqref{heisenb}, one obtains the two following  equations involving only real quantities, 
\begin{equation}
\label{pqeveq}
  \hbar \frac{\partial}{\partial t}|\psi(t)\rg= \widetilde H(t)\, |\psi(t)\rg\, , \quad \hbar\frac{\ud  A_H}{\ud t} = [A_{H},\widetilde H] + U(t,t_0)^{\dag}\left(\hbar\frac{\partial A}{\partial t}\right) U(t,t_0)\, , 
\end{equation}
where $U(t,t_0)$ is the rotation \eqref{evrot}.
They are our \textit{Baby Majorana equations} \cite{wilczek09}.
The price to pay with our manipulation is that the eigenvalues $\pm \ii \mcE$ of $\widetilde H$ are real only in the trivial case $\mcE=0$: no physical energy comes out from our formalism. 

If $\mcE$ is constant, the pseudo-Hamiltonian $\widetilde H$ generates uniform rotational evolution in the plane, i.e., $U(t,t_0)= e^{\frac{\mcE}{\hbar} t \tau_2}$.  
In the case of the time evolution of a pure state $P_{\phi_0}$, this evolution corresponds to  a   uniform rotation, with frequency $\vert \mcE\vert/\hbar$, of the unit vector $|\phi_0\rg$.

On the algebraic side, we note that our real quantum dynamics involves the antisymmetric matrix $\tau_2$. Together with $\sigma_1$ and $\sigma_3$ these 3 matrices generate the Lie algebra $\mathfrak{sl}(2,\R)$ of the group SL$(2,\R)$ 
\begin{equation}
\label{sl2R}
[\sigma_1,\tau_2]= 2\sigma_3\, , \quad [\tau_2,\sigma_3]= 2\sigma_1\, , \quad [\sigma_3,\sigma_1]= -2\tau_2\,. 
\end{equation}

If we had  sticked to the complex formalism imposed by \eqref{eveq}, we would have to introduce 
 the third Pauli matrix $\sigma_2$, which was missing in \eqref{Pauli13}. Then   the self-adjointness of $H$ is preserved, and quantum evolution restores the full real Lie algebra of SU$(2)$ generated  by the $\sigma_i$'s, $i=1,2,3$. Actually, we will use  in Section \ref{magnint} the equation \eqref{eveq}  for presenting  a physical example of \eqref{hamsol}.

\subsection{Open systems}
Restricting the form of ``acceptable'' Hamiltonians to \eqref{hamsol1} or \eqref{pseudoH}  is not tenable  since this family does not include the case of quantum \textit{open} systems, particularly when the latter are submitted to measurement. We will come back to the important question of measurement  in Section \ref{quantmeas} with the presentation of a simple example. In this subsection, we consider an extension of  \eqref{pqeveq} to a type of \textit{quantum master}   equation, namely the Lindblad equation \cite{breupetr02},  describing the time evolution of the density matrix $\rho_{r,\phi}$ in \eqref{standrhomain} for an open system. The interest for a such an equation has been recently revived by S. Weinberg in a large audience article \cite{weinberg17}. When applied to the evolution of a quantum observable $A$, the Lindblad equation in its diagonal form (which does not restrict its validity)  is expressed as 
\begin{equation}
\label{lindbladA}
\frac{\ud A}{\ud t}=\frac{\ii}{\hbar}\,[H,A]+\sum _{k}h_k\left[L_{k}AL_{k}^{\dagger }-\frac{1}{2}\left(AL_{k}^{\dagger}L_{k}+L_{k}^{\dagger}L_{k}A\right)\right]\, ,
\end{equation}
where the  $L_k$ with the identity form a basis of operators and the coefficients $h_k$ are non-negative constants. 

In a finite $N$-dimensional Hilbert space there are $N^2 -1$  Lindblad operators $L_k$ that,  together with the identity, generate the entire $N^2$-matrix vector space. 
Since in the present  framework 
our Hilbert space is two-dimensional ($N=2$), 
the three matrices $\sigma_1$, $\tau_2$, $\sigma_3$, together with the identity, form a basis for the vector space of  real $2\times 2$ matrices, the equation \eqref{lindbladA}, when  applied to $\rho_{r,\phi}$, becomes
 \begin{equation}
\label{lindbladrho}
\begin{split}
\frac{\ud \rho_{r,\phi}}{\ud t}&=\frac{1}{\hbar}\,[\rho_{r,\phi}, \widetilde{H}]+ h_1 \left(\sigma_1\rho_{r,\phi}\sigma_1- \rho_{r,\phi}\right) + h_2 \left(-\tau_2\rho_{r,\phi}\tau_2 +\rho_{r,\phi}\right) + h_3 \left(\sigma_3\rho_{r,\phi}\sigma_3- \rho_{r,\phi}\right)\\
& = r\left(\frac{\mathcal{E}(t)}{\hbar}\,\sin2\phi - h_1\,\cos2\phi\right)\,\sigma_3 + r\left(-\frac{\mathcal{E}(t)}{\hbar}\,\cos2\phi - h_3\,\sin2\phi\right)\,\sigma_1 + h_2\,I\,. 
\end{split}
\end{equation}
Since $\dfrac{\ud \rho_{r,\phi}}{\ud t}\equiv \dot  \rho_{r,\phi}= \frac{1}{2}(\dot r \cos2\phi -2r \sin2\phi \,\dot\phi) \sigma_3 + \frac{1}{2}(\dot r \sin2\phi +2r \cos2\phi\,\dot\phi) \sigma_1$, we get by identification and after some algebraic manipulations using \eqref{sl2R} the condition 
\begin{equation}
\label{linddiff3}  h_2=0\,, 
\end{equation}
and
 the first-order differential system
\begin{align}
\label{linddiff1}
   \dot \phi &= \frac{h_1-h_3}{2} \sin4\phi -\frac{\mathcal{E}}{\hbar}\, ,   \\
 \label{linddiff2}  \frac{\dot r}{r} &= (h_3-h_1) \,\cos4\phi - (h_3 + h_1)\, .
\end{align}
Eq.\;\eqref{linddiff1} leads to a kind of Ricatti equation. Having in hand a solution $\phi(t)$,  the second one is easily integrated and we find
\begin{equation}
\label{sol}
r(t)= r_0 \exp\left[- (h_3 + h_1)(t-t_0) + (h_3-h_1)\int_{t_0}^{t}\cos4\phi(t^{\prime})\, \ud t^{\prime} \right]\, . 
\end{equation}
Due to the nonnegativess of $h_1$ and $h_3$, we easily infer that $r\to 0$, i.e.,  $\rho_{r,\phi}\to I/2$ as $t\to \infty$, which means that the von Neumann entropy of the open system tends to its maximum at large time, as we can expect.  Note that for $h_1=h_3$ the angle $\phi$ behaves 
like for a closed system, 
\begin{equation}
\label{h1eqh3a}
\phi(t)-\phi(t_0)= -\int_{t_0}^{t}\ud t^{\prime}\, \frac{\mathcal{E}(t^{\prime})}{\hbar}\, , 
\end{equation}
whereas the behaviour of $r$ simplifies to
\begin{equation}
\label{h1eqh3a}
r(t)= r_0 e^{- 2h_1 (t-t_0)}\, . 
\end{equation}

Finally, it is interesting to  establish  the semi-classical portrait of Eq.\;\eqref{lindbladrho} with $h_2=0$  by calculating with lower symbols formula  \eqref{symsigma} and \eqref{standrhomainlus} its mean value in state $|\theta\rg$   
\begin{equation}
\label{dynclass}
\begin{split}
\frac{\ud}{\ud t}\lg \theta|\rho_{r,\phi}|\theta\rg&=\frac{ \dot r}{2}\cos2(\phi-\theta) - r\sin2(\phi-\theta) \dot\phi\\
&= r\frac{\mathcal{E}}{\hbar}\sin2(\phi-\theta) -rh_1\cos2\phi\cos2\theta-rh_3\sin2\phi\sin2\theta\,. 
\end{split}
\end{equation}
Since such an equation should hold for all $\theta\in [0,2\pi)$, it leads to the same differential system as the quantum \eqref{linddiff1}-\eqref{linddiff2}.
This is a simple illustration of the Ehrenfest theorem (see \cite{weinberg98} and references therein), which is exact here.  
Similarly, by using again the formulas \eqref{symsigma} corresponding to the quantization \eqref{quantcircle13} with pure states,  we  check that \eqref{lindbladrho} is the quantization of the classical dynamical system for the time dependent parameters $(r,\phi)$, 
\begin{equation}
\label{classdyn}
\begin{split}
&\frac{\ud}{\ud t} \widehat{\rho}_{r,\phi}(\theta) \\ &=2 r\left[\left(\frac{\mathcal{E}(t)}{\hbar}\,\sin2\phi - h_1\,\cos2\phi\right)\cos 2\theta + \left(-\frac{\mathcal{E}(t)}{\hbar}\,\cos2\phi - h_3\,\sin2\phi\right)\sin2\theta\right]\, ,
\end{split}
\end{equation}
which is the same as the quantum \eqref{linddiff1}-\eqref{linddiff2}. 
\section{Entanglement, Baby Bell inequality, and isomorphisms}
\label{entanglement}

\subsection{Bell states}
The possible entanglement property of quantum states appears with the construction of tensor products \cite{tensor} of Hilbert spaces for describing quantum states of composite systems. In the present case, we are in presence of a remarkable sequence of vector space isomorphisms  due to the fact that $2^2 = 2\times 2 = 2+2$: 
\begin{equation}
\label{seqiso}
\R^2 \otimes \R^2 \cong \R^2\times \R^2 \cong \R^2 \oplus \R^2  \cong \C^2\cong \H\, , 
\end{equation}
where $\H$ is the field of quaternions (see Appendix  \ref{allSU2} for definition and properties of these objects). 

Let us  first write  the (canonical) orthonormal basis of the tensor product $\R_A^2 \otimes \R_B^2$, the first factor being for system ``$A$'' and the other for system ``$B$'',  as
\begin{equation}
\label{tensbas}
|0\rg_A\otimes|0\rg_B\, , \quad \left|\frac{\pi}{2}\right\rg_A\otimes\left|\frac{\pi}{2}\right\rg_B\, , \quad |0\rg_A\otimes\left|\frac{\pi}{2}\right\rg_B\, , \quad \left|\frac{\pi}{2}\right\rg_A\otimes|0\rg_B\, .
\end{equation}
These factors $|0\rg$, $\left|\frac{\pi}{2}\right\rg$ pertaining to $A$ or to $B$,  are termed ``$q$-bit'' or ``qubit" in standard language of  quantum information. Actually, it would be preferable to call them ``Rebit" since this term exists as well in quantum information when one manipulates states in real Hilbert spaces (see for instance \cite{caves_etal01}.   They can be associated to an orientation pointer measuring the horizontal (resp. vertical)  direction or polarisation described by the state $|0\rg$  (resp. $\left|\frac{\pi}{2}\right\rg$). 

We now introduce the celebrated Bell pure states \cite{nielsen-chuang00,rieffel_polak11} in $\R_A^2 \otimes \R_B^2$ as special linear superpositions of the above basis elements
\begin{align}
\label{bell1}
  |\Phi^{\pm} \rg & = \frac{1}{\sqrt 2} \left( |0\rg_A\otimes|0\rg_B \pm \left|\frac{\pi}{2}\right\rg_A\otimes\left|\frac{\pi}{2}\right\rg_B\right) \, , \\
 \label{bell2}   |\Psi^{\pm} \rg&  = \frac{1}{\sqrt 2} \left(  |0\rg_A\otimes\left|\frac{\pi}{2}\right\rg_B \pm \left|\frac{\pi}{2}\right\rg_A\otimes|0\rg_B  \right)\,,
\end{align}
or in  matrix form,

\begin{align}
\label{Bellmat}
\nonumber &\begin{pmatrix}
   |\Phi^{+} \rg    &    |\Phi^{-} \rg&  |\Psi^{+} \rg&  |\Psi^{-} \rg  
\end{pmatrix} \\ &= \begin{pmatrix}
  |0\rg_A\otimes|0\rg_B & \left|\frac{\pi}{2}\right\rg_A\otimes\left|\frac{\pi}{2}\right\rg_B & |0\rg_A\otimes\left|\frac{\pi}{2}\right\rg_B & \left|\frac{\pi}{2}\right\rg_A\otimes|0\rg_B
\end{pmatrix} \frac{1}{\sqrt{2}}\begin{pmatrix}
  1    &  1 & 0 & 0  \\
   1   &  -1 & 0& 0\\
   0 & 0 & 1 & 1\\
   0 & 0 & 1 & -1
\end{pmatrix}\, .
\end{align}
They form an orthonormal basis of $\R_A^2 \otimes \R_B^2$. They are four specific \emph{maximally entangled} quantum states of two qubits. ``Maximally'' refers to the degree to which a state in a quantum system consisting of two entities is entangled. This degree is measured by the aforementioned von Neumann entropy \cite{VNentropy}  of either of the two reduced density operators of the state, i.e., obtained by tracing out one of the partners in the composite systems. They exhibit perfect correlation. 
Consider for instance the state $|\Phi^{+} \rg$. If the pointer associated to $A$ measures its qubit in the standard basis the outcome would be perfectly random, either possibility having probability 1/2. But if the pointer associated to $B$ then measures its qubit, the outcome, although random for it alone,   is the same as the one $A$ gets. There is quantum correlation. Tracing out the $B$-partner in the pure state $ |\Phi^{+} \rg\lg  \Phi^{+} |$ gives the mixed state $\frac{1}{2}I_A$ for the $A$-partner, whose the von Neumann entropy  \eqref{VNentr} is equal to its maximal value $\ln 2$. 

\subsection{Inequality and its violation}
Let us transpose into the present setting the  1964 analysis and result \cite{bell64} presented by Bell  in his discussion about the EPR paper \cite{epr35} and about the subsequent Bohm's  approaches \cite{bohm52,bohm_ah57} based on the assumption  of  hidden variables. We replace the spin one-half particles considered by Bell as examples with the parallel (i.e., $+1$) and perpendicular (i.e., $-1$) quantum orientations in the plane as the only possible issues of the observable $\sigma_{\phi}$ introduced in \eqref{sigphi}, supposing that there exists a  pointer device designed for measuring such orientations with outcomes $\pm 1$ only. We  now consider a bipartite system of two quantum orientations described respectivement by  $\overrightarrow{\boldsymbol{\sigma}}^A$ and $\overrightarrow{\boldsymbol{\sigma}}^B$ in the so-called \textit{singlet} state $\Psi^{-}$ in \eqref{bell1}. In such a state, if measurement of the component $\sigma^A_{\phi_a}:=\overrightarrow{\boldsymbol{\sigma}}^A\cdot \widehat{\mathbf{u}}_{\phi_a}$ yields the value $+1$, then measurement of $\sigma^B_{\phi_b}$ when $\phi_b= \phi_a$  must yield the value $-1$, and vice-versa. From a classical perspective, the explanation of such a correlation needs a predetermination by means of the existence of \textit{hidden} parameters $\lambda\in \Lambda$, with  the notations of \cite{bell64}. The result $\varepsilon^A\in \{-1,+1\}$ (resp. $\varepsilon^B\in \{-1,+1\}$)  of measuring $\sigma^A_{\phi_a}$ (resp. $\sigma^B_{\phi_b}$ )  is then determined by $\phi_a$ \underline{and} $\lambda$ only, \underline{not} by $\phi_b$ also, i.e. $\varepsilon^A= \varepsilon^A(\phi_a ,\lambda)$ (resp. $\varepsilon^B= \varepsilon^B(\phi_b ,\lambda)$). Given a probability distribution $\rho(\lambda)$ on $\Lambda$, the \underline{classical} expectation value of the product of the two components $\sigma^A_{\phi_a}$ and $\sigma^B_{\phi_b}$ is given by
\begin{equation}
\label{classexpec}
\mathsf{P}(\phi_a,\phi_b)= \int_{\Lambda}\ud \lambda\, \rho(\lambda) \, \varepsilon^A(\phi_a ,\lambda)\,\varepsilon^B(\phi_b ,\lambda)
\, . 
\end{equation}
Since $ \int_{\Lambda}\ud \lambda\, \rho(\lambda)= 1$ and $\varepsilon^{A,B}= \pm1$, we have $-1 \leq \mathsf{P}(\phi_a,\phi_b)\leq 1$.
Equivalent predictions with the quantum setting then imposes the equality between the classical and quantum expectation values:
\begin{equation}
\label{equalexpec}
\mathsf{P}(\phi_a,\phi_b)= \left\lg \Psi^{-} \right| \sigma^A_{\phi_a}\otimes \sigma^B_{\phi_b}\left|  \Psi^{-} \right\rg = - \widehat{\mathbf{u}}_{\phi_a}\cdot \widehat{\mathbf{u}}_{\phi_b}= -\cos(\phi_a-\phi_b) \,. 
\end{equation}
In \eqref{equalexpec},  the value $-1$ is reached at $\phi_a = \phi_b$. This is possible for $\mathsf{P}(\phi_a,\phi_a)$ only if $\varepsilon^A(\phi_a ,\lambda) = -\varepsilon^B(\phi_a ,\lambda)$. Hence,  we can write \eqref{classexpec} as
\begin{equation}
\label{classexpect1}
\mathsf{P}(\phi_a,\phi_b)= -\int_{\Lambda}\ud \lambda\, \rho(\lambda) \, \varepsilon(\phi_a ,\lambda)\,\varepsilon(\phi_b ,\lambda)\, , \quad  \varepsilon(\phi,\lambda)\equiv  \varepsilon^A(\phi,\lambda) = \pm1
\, . 
\end{equation}
Let us now introduce a third unit vector $\widehat{\mathbf{u}}_{\phi_c}$.  Due to $\varepsilon^2= 1$, we have
\begin{equation}
\label{Pabc}
\mathsf{P}(\phi_a,\phi_b) - \mathsf{P}(\phi_a,\phi_c)= \int_{\Lambda}\ud \lambda\, \rho(\lambda) \, \varepsilon(\phi_a ,\lambda)\,\varepsilon(\phi_b ,\lambda)\, \left[\varepsilon(\phi_b ,\lambda)\,\varepsilon(\phi_c ,\lambda)-1\right]\, . 
\end{equation}
It results the (baby) Bell inequality:
\begin{equation}
\label{Pabcinf}
\vert\mathsf{P}(\phi_a,\phi_b) - \mathsf{P}(\phi_a,\phi_c)\vert\leq  \int_{\Lambda}\ud \lambda\, \rho(\lambda)\, \left[1-  \varepsilon(\phi_b ,\lambda)\,\varepsilon(\phi_c ,\lambda)\right]= 1+ \mathsf{P}(\phi_b,\phi_c)\, . 
\end{equation}
Hence, the validity of the existence of hidden variable(s) for justifying the quantum correlation  in the singlet state $\Psi^{-}$, and which is encapsulated by \eqref{Pabcinf},  has the following consequence on the arbitrary triple $(\phi_a,\phi_b,\phi_c)$:
\begin{equation}
\label{ineqcos}
1-\cos(\phi_b-\phi_c) \geq \left\vert \cos(\phi_b-\phi_a) - \cos(\phi_c-\phi_a)\right\vert\, , 
\end{equation} 
equivalently, in terms of the two independent angles $\zeta= \dfrac{\phi_a-\phi_b}{2}$ and  $\eta= \dfrac{\phi_b-\phi_c}{2}$,
\begin{equation}
\label{ineqsin}
\left\vert \sin^2 \zeta - \sin^2(\eta+\zeta)\right\vert \leq \sin^2\eta\, . 
\end{equation}
It is easy to find pairs $(\zeta, \eta)$ for which this inequality does not hold true. For instance with $\eta=\zeta \neq 0$, i.e., $\phi_b= \dfrac{\phi_a + \phi_c}{2}$, \eqref{ineqsin} becomes 
 \begin{equation}
\label{ineqsin1}
\vert 4\sin^2\eta - 3\vert \leq 1\, , 
\end{equation}
and this does not hold true for all $\vert \eta\vert < \pi/4$, i.e., for $\left\vert \phi_a-\phi_b\right\vert = \left\vert \phi_b-\phi_c\right\vert< \pi/2$. 
 
Actually, we did not follow  the proof given by Bell in \cite{bell64}, which is a lot more elaborate. Also, Bell considered  unit vectors in $3$-space. Restricting his proof to vectors in the plane does not make any difference, as it is actually the case in many works devoted to the foundations of quantum mechanics, see for instance \cite{harrspek10}. 

\subsection{Complex cat states}
Let us now turn our attention to the complex two-dimensional Hilbert space $\C^2$, one of the most familiar mathematical objects in our initiation to quantum formalism, 
after having been so nicely  popularised by   Feynman \cite{feynman3} with examples like two-level atom or spin one-half objects. Now, as a complex vector space, space $\C^2$, with canonical basis $\mathbf{e}_1$, $\mathbf{e}_2$, has a real structure. The latter is isomorphic to the  real vector space  $\R^4$, itself isomorphic to $\R^2 \otimes \R^2$, from the above. A standard real structure is obtained by considering the vector expansion
\begin{equation}
\label{}
\C^2 \in \mathbf{v} = z_1 \mathbf{e}_1 + z_2 \mathbf{e}_2 = x_1 \mathbf{e}_1 + y_1 \left(\ii \mathbf{e}_1\right) +  x_2 \mathbf{e}_2 + y_2 \left(\ii \mathbf{e}_2\right)\, , 
\end{equation}
i.e., by writing $z_1= x_1 + \ii y_1$, $z_2= x_2 + \ii y_2$, and considering  the set of vectors 
\begin{equation}
\label{e1e2i}
\left\{\mathbf{e}_1,   \mathbf{e}_2, \left(\ii \mathbf{e}_1\right), \left(\ii \mathbf{e}_2\right)\right\} 
\end{equation}
as forming a  basis of $\R^4$. Now, we need to define a map from this $\R^4$ to $\R^2 \otimes \R^2$, forgetting about the superfluous subscripts $A$ and $B$. For this, we use the map [Euclidean plane $\R^2$] $\mapsto$ [complex ``plane''  $\C$]  determined by
 \begin{equation}
\label{R2C}
|0\rg \mapsto 1\, , \qquad \left|\frac{\pi}{2}\right\rg \mapsto \ii\, . 
\end{equation}
Considering this identification as a guidance, we choose to write the correspondence between bases as
\begin{equation}
\label{R4C2bas}
  |0\rg\otimes|0\rg = \mathbf{e}_1\, , \quad \left|\frac{\pi}{2}\right\rg\otimes\left|\frac{\pi}{2}\right\rg= - \mathbf{e}_2\, , \quad 
  |0\rg\otimes\left|\frac{\pi}{2}\right\rg = \left(\ii \mathbf{e}_1\right) \, , \quad  \left|\frac{\pi}{2}\right\rg\otimes|0\rg = \left(\ii \mathbf{e}_2\right) \, ,
\end{equation}
or, in matrix form, 
\begin{align}
\label{R4C2}
\nonumber &\begin{pmatrix}
  |0\rg\otimes|0\rg & \left|\frac{\pi}{2}\right\rg\otimes\left|\frac{\pi}{2}\right\rg & |0\rg\otimes\left|\frac{\pi}{2}\right\rg & \left|\frac{\pi}{2}\right\rg\otimes|0\rg
\end{pmatrix}\\ &=    \begin{pmatrix}   \mathbf{e}_1&   \mathbf{e}_2 & \left(\ii \mathbf{e}_1\right) & \left(\ii \mathbf{e}_2\right) 
\end{pmatrix} \begin{pmatrix}
  1    &  0 & 0 & 0  \\
   0  &  -1 & 0& 0\\
   0 & 0 & 0 & 1\\
   0 & 0 & 1 & 0
\end{pmatrix}\, .
\end{align}
Interpreting $\C^2$ as the space of quantum states of a two-level system, e.g. spin one-half with  standard arrow notations
\begin{equation}
\label{spinupdown}
\mathbf{e}_1 := \upa\equiv \begin{pmatrix}
      1   \\
      0  
\end{pmatrix}\, , \qquad    \mathbf{e}_2 \equiv \dwa :=\begin{pmatrix}
      0    \\
      1  
\end{pmatrix}\, ,  
\end{equation} 
the  correspondences \eqref{R4C2bas} and \eqref{R4C2} might be interpreted as giving a real composite nature to ``up'' and ``down" complex objects. 
Note that the $4\times4$-matrix on the right of \eqref{R4C2} is its own inverse. Finally, we get the matrix map from the Bell basis to the basis \eqref{e1e2i} 
\begin{equation}
\label{Bellmate}
\begin{pmatrix}
   |\Phi^{+} \rg    &    |\Phi^{-} \rg&  |\Psi^{+} \rg&  |\Psi^{-} \rg  
\end{pmatrix} = \begin{pmatrix}
   \mathbf{e}_1&   \mathbf{e}_2 & \left(\ii \mathbf{e}_1\right) & \left(\ii \mathbf{e}_2\right) 
\end{pmatrix} \frac{1}{\sqrt{2}}\begin{pmatrix}
  1    & 1 & 0 & 0  \\
  - 1   &  1 & 0& 0\\
   0 & 0 & 1 &- 1\\
   0 & 0 & 1 & 1
\end{pmatrix}\, .
\end{equation}
 In terms of components of vectors in their respective spaces, 
 \begin{equation}
\label{Bellmatcomp}
\begin{pmatrix}
  x_1    \\    x_2 \\  y_1 \\  y_2 
\end{pmatrix} = \frac{1}{\sqrt{2}}\begin{pmatrix}
  1    & 1 & 0 & 0  \\
  - 1   &  1 & 0& 0\\
   0 & 0 & 1 & -1\\
   0 & 0 & 1 & 1
\end{pmatrix}  \begin{pmatrix}
   x^+\\  x^- \\ y^+ \\ y^-
\end{pmatrix}\, . 
\end{equation}
Equivalently, in complex notations, with $z^\pm = x^\pm + \ii y^\pm$, 
\begin{equation}
\label{BellC2}
\begin{pmatrix}
      z^+    \\
      z^-  
\end{pmatrix}= \frac{1}{\sqrt{2}}\begin{pmatrix}
    1  &  -\mathsf{C}  \\
    \mathsf{C}  & 1  
\end{pmatrix}\begin{pmatrix}
      z_1    \\
      z_2  
\end{pmatrix}\equiv \mathcal{C}_{@} \begin{pmatrix}
      z_1    \\
      z_2  
\end{pmatrix}\, , 
\end{equation}
where we have introduced the conjugation  operator $\mathsf{C} z  = \bar z$, i.e. the mirror symmetry with respect to the real axis, $-\mathsf{C}$ being the  mirror symmetry with respect to the imaginary  axis. Note that the ``cat'' operator $\mathcal{C}_{@}$ can be expressed in the quaternionic language for SU$(2)$, as introduced in \eqref{corSU2H}, by
\begin{equation}
\label{catSU2}
\mathcal{C}_{@} =\frac{1}{\sqrt{2}}\left( I + \mathsf{F}\right)\, , \quad \mathsf{F} :=C\, \widehat{\boldsymbol{\jmath}}= \begin{pmatrix}
   0   &  -C  \\
  C    &  0
\end{pmatrix}\, , \quad \widehat{\boldsymbol{\jmath}} =-\ii \sigma_2= \begin{pmatrix}
  0    &  -1  \\
   1 &  0
\end{pmatrix}\, .  
\end{equation}
Therefore, with the above choice of isomorphisms, Bell entanglement in $\R^2\otimes \R^2$ is \underline{not} represented by a linear superposition in $\C^2$. It involves also the two mirror symmetries $\pm \sfC$. Interpreting $\C^2$ as in \eqref{spinupdown}, 
we see that the  operator $\mathsf{F}$ is a kind of ``flip'' whereas  the operator $\mathcal{C}_{@}$  builds from the  \textit{up} and \textit{down} basic states the two elementary Schr\"odinger cats
\begin{equation}
\label{Scats}
\begin{split}
\mathsf{F}\,\upa &= \dwa\, , \quad \mathsf{F}\,\dwa = -\upa\, , \\ \mathcal{C}_{@}\,\upa &= \frac{1}{\sqrt{2}}(\upa + \dwa)\, , \quad \mathcal{C}_{@}\,\dwa= \frac{1}{\sqrt{2}}(-\upa + \dwa)\, . 
\end{split}
\end{equation}
 We can now understand better the real Bell state nature of these elementary cat superpositions if we remember the identification \eqref{R4C2bas} between composite real states and complex \textit{up} and \textit{down} basic states.
Interesting too is the appearance of the flip $\mathsf{F}$ in the construction of  the spin one-half  coherent state defined in terms of spherical coordinates $(\theta,\phi)$ by
\begin{equation}
\label{spinCS}
\mathbb{S}^2 \ni \widehat{\mathbf{n}}(\theta,\phi) \mapsto |\theta,\phi \rg =\left(\cos\frac{\theta}{2}\,\upa + e^{\ii \phi}\sin\frac{\theta}{2}\,\dwa\right)\equiv \begin{pmatrix}
     \cos\frac{\theta}{2}     \\
   e^{\ii \phi}\sin\frac{\theta}{2}     
\end{pmatrix}\, . 
\end{equation}
Indeed, they result from the action on the up state by the  operator  $D^{\frac{1}{2}}\left(\bar\xi_{\widehat{\mathbf{n}}}\right)$ given in \eqref{D12bxi}. This operator represents the element  $\bar\xi_{\widehat{\mathbf{n}}}$ in SU$(2)$,  quaternionic conjugate or inverse of  $\xi_{\widehat{\mathbf{n}}}$. The latter  corresponds through the homomorphism SO$(3)$ $\mapsto$ SU$(2)$ to  the specific rotation $\mathcal{R}_{\widehat{\mathbf{n}}}$ that
maps the unit vector pointing to the north pole, $\widehat{\mbox{\textbf{\textit{k}}}}=(0,0,1)$,
to $\widehat{\mathbf{n}}$, as is shown in Figure \ref{rotkton} and explained  at length in Appendix \ref{allSU2}:
\begin{equation}
\label{Scatscs}
 |\theta,\phi \rg = D^{\frac{1}{2}}\left(\bar\xi_{\widehat{\mathbf{n}}}\right)\upa =  \begin{pmatrix}
   \cos\frac{\theta}{2}   &- \sin\frac{\theta}{2} e^{-\ii \phi}  \\
  \sin\frac{\theta}{2} e^{\ii \phi}   &  \cos\frac{\theta}{2}
\end{pmatrix}\begin{pmatrix}
      1   \\
      0 
\end{pmatrix}\, . 
\end{equation}
Now we notice that the second column of $D^{\frac{1}{2}}\left(\bar\xi_{\widehat{\mathbf{n}}}\right)$ is precisely the flip of the first one, 
\begin{equation}
\label{flipcol}
D^{\frac{1}{2}}\left(\bar\xi_{\widehat{\mathbf{n}}}\right) = \begin{pmatrix}
    |\theta,\phi \rg &     \mathsf{F} |\theta,\varphi \rg\end{pmatrix}\,. 
\end{equation}
Actually, this feature is the key for grasping the isomorphisms $\C^2 \cong \H \cong \R_+ \times \mathrm{SU}(2)$. With the notations of Appendix \ref{allSU2} and using quaternionic algebra, e.g.,  $\widehat{\boldsymbol{\imath}}= \widehat{\boldsymbol{\jmath}} \widehat{\mbox{\textbf{\textit{k}}}}$ (and even permutations),  
\begin{equation}
\label{C2H}
\H \ni q = q_0 + q_1 \widehat{\boldsymbol{\imath}} + q_2 \widehat{\boldsymbol{\jmath}} + q_3 \widehat{\mbox{\textbf{\textit{k}}}} = 
q_0  + q_3 \widehat{\mbox{\textbf{\textit{k}}}}  + \widehat{\boldsymbol{\jmath}} \left(q_1 \widehat{\mbox{\textbf{\textit{k}}}}  + q_2 \right)\equiv \begin{pmatrix}
   q_0  + \ii q_3       \\
     q_2 + \ii q_1   
\end{pmatrix} \equiv Z_q \in \C^2\, , 
\end{equation}
after identifying  $ \widehat{\mbox{\textbf{\textit{k}}}}\equiv \ii$, as both are roots of $-1$.  Then the flip appears naturally in the final identification $\H \cong \R_+ \times \mathrm{SU}(2)$ as
\begin{equation}
\label{HRSU2 }
q \equiv \begin{pmatrix}
 q_0  + \ii q_3     & - q_2 + \ii q_1   \\
 q_2 + \ii q_1     &  q_0  - \ii q_3
\end{pmatrix} = \begin{pmatrix}
 Z_q     &  \mathsf{F} Z_q  
\end{pmatrix}\,. 
\end{equation}

\section{Magnetic interpretation}
\label{magnint}
The fact that the sole acceptable form the Hamiltonian may assume in our model is  \eqref{hamsol1} might suggest a  magnetic field interacting with a one-half spin magnetic moment. How can we integrate this physics in our toy model? 

We recall that the (potential) energy of a magnetic moment \cite{feynman2} $\vec{\boldsymbol{\mu}}$  submitted to a magnetic field $\overrightarrow{\mathbf{B}}$ is given by 
\begin{equation}
\label{magen}
V_m = - \overrightarrow{\mathbf{B}}\cdot \vec{\boldsymbol{\mu}}\, . 
\end{equation}
In the case of a moving  charged point, the magnetic moment is proportional to the angular momentum of this charge. Here we consider a magnetic moment as a physical system on its own, without any consideration of space location or something else, and proportional  to an angular momentum or classical spin denoted by $\vec{\mathbf{J}}$:
 \begin{equation}
\label{gyro}
\vec{\boldsymbol{\mu}} = \gamma  \vec{\mathbf{J}}\, . 
\end{equation}
The factor $\gamma$ is usually named gyromagnetic factor. 
Now, the energy of a magnetic moment submitted to a constant magnetic field alone is conserved. Choosing the  direction of the unit vector $\widehat{\mbox{\textbf{\textit{k}}}}$, part of the orthonormal basis in space $\left(\widehat{\boldsymbol{\imath}}, \widehat{\boldsymbol{\jmath}},\widehat{\mbox{\textbf{\textit{k}}}}\right)$,  as the same as $\overrightarrow{\mathbf{B}}$, we have for the Hamiltonian
\begin{equation}
\label{classhamilt}
h = V_m= - \gamma\Vert \overrightarrow{\mathbf{ B}}\Vert \,\Vert \vec{\mathbf{ J}} \Vert \cos\theta = \gamma B J \cos\theta= E,
\end{equation} 
and so the polar angle $\theta$ is conserved: the magnetic moment experiences a precession about the $\widehat{\mbox{\textbf{\textit{k}}}}$-axis at constant azimuthal velocity $\dot\phi$ in such a way that $J_z = J \cos\theta$ is also conserved.  Actually, this precession motion which takes place on the surface of a sphere $\mathbb{S}^2(J)$ of radius $J$ has to be viewed as a phase trajectory in $\mathbb{S}^2(J)$, the latter being  \underline{viewed} as the classical phase space for the magnetic moment. The canonical conjugate coordinates are in the present case action-angle variables, the  angle variable being the azimuthal angle $\phi$ whereas the action variable $I_{\phi}$ is precisely $J_3$. Indeed, from the expression 
\begin{equation}
\label{2form}
J \, \sin\theta\, \ud\theta \, \ud\phi = \ud\phi\, \ud J_z
\end{equation} of the rotational invariant measure on  $\mathbb{S}^2(J)$,  one infers that $J_z$ is canonically conjugate to $\phi$, and the action variable defined as  
\begin{equation}
\label{actvar}
I_{\phi} = \frac{1}{2\pi}\oint J_z\, \ud\phi 
\end{equation}
is identical to $J_z$ in the case of a constant external magnetic field.
Note that $J_z$ and $\phi$ are  a part of the  cylindrical coordinates of the sphere $\mathbb{S}^2(J)$ with radius $J$.

Let us turn our attention to the quantum model of a magnetic moment interacting with a constant magnetic field. Following the integral quantization procedure of the two-sphere $\mathbb{S}^2$ and of functions (or distributions on it),  as is explained in details in Appendix \ref{unitsphere},  and yielding the expression \eqref{ntosig}, the quantum version of the classical Hamiltonian \eqref{magen} with $ \vec{\mathbf{J}}= J \widehat{\mathbf{n}}$ reads as
\begin{equation}
\label{quanthm}
H= A_h = -\frac{r}{3}\gamma J \vec{\mathbf{B}}\cdot \vec{\boldsymbol{\sigma}}\, ,  
\end{equation}
where the parameter $0\leq r \leq 1$ plays the same r\^ole as in \eqref{standrhomain}. 
Now, contrarily to our previous choice, we take the direction of the magnetic field along the unit vector $\widehat{\boldsymbol{\jmath}}$,  $\overrightarrow{\mathbf{B}}= B \widehat{\boldsymbol{\jmath}}$. This new choice gives precisely the quantum Hamiltonian  \eqref{hamsol1},
\begin{equation}
\label{buphi}
H =  \begin{pmatrix}
   0  & -\ii \mcE   \\
   \ii  \mcE  &  0
\end{pmatrix}\, , \quad \mbox{with} \quad \mcE = - \frac{r}{3}\gamma J B\,. 
\end{equation}
This formula leaves to us the freedom to choose $r$ and $J$ in order to comply with observations, for instance for the electron. We just remind that, in its quantum  description,  the electron \textit{total} spin magnetic moment is given by the formula \cite{resnick_eisberg85}:
\begin{equation}
\label{mafmom}
\Vert \vec{\boldsymbol{\mu}}_s\Vert = -\frac{e}{2m}g\Vert \vec{\mathbf{J}}\Vert\, , \quad \vert \vec J\Vert = \sqrt{\frac{1}{2}\left(\frac{1}{2}+1\right)}\hbar= \frac{\sqrt{3}}{2} \hbar\, , 
\end{equation}
where
$e$ is the charge of the electron, $g$ is the Land\'e $g$-factor, and by the quantization equation along the  direction determined by $\widehat{\mathbf{n}}$:
\begin{equation}
\label{mun}
\mu_{\widehat{\mathbf{n}}}= \pm \frac{1}{2}g{\mu_B}
\end{equation}
where $\mu_B = e\hbar/2m_{e}$ is the Bohr magneton.

\section{An example of  (baby) quantum measurement}
\label{quantmeas}

In this section we adapt to  the set of orientations in the Euclidean plane an example of simple quantum measurement presented by Peres in \cite{peres90}. Two planes and their tensor product are under consideration here. The first one is the space of states $|\phi_M\rg$ of an  orientation \textit{pointer}. We note that the action of the matrix  $\tau_2$ on these states corresponds to the rotation by $\pi/2$:
\begin{equation}
\label{tau2act}
\tau_2\,|\phi_M\rg = \begin{pmatrix}
   0   & -1   \\
   1   &  0
\end{pmatrix}\, \begin{pmatrix}
      \cos\phi_M    \\
      \sin\phi_M 
\end{pmatrix} = \begin{pmatrix}
       - \sin\phi_M    \\
         \cos\phi_M
\end{pmatrix} = \left|\phi_M + \frac{\pi}{2}\right\rg\,. 
\end{equation}
The second plane is the space of states $|\phi_S\rg$ of the system whose the orientation is to be measured. 
Let us now introduce the quantum observable  $A^S$ to be measured. It acts  on the states of the system and with spectral decomposition like in \eqref{spmeasR2},
\begin{equation}
\label{pointer}
A^S= \lambda_{\parallel} \, P_{\phi_{\parallel}} + \lambda_{\perp} \, P_{\phi_{\perp}}\,, \quad \phi_{\perp}= \phi_{\parallel} \pm \frac{\pi}{2}\, . 
\end{equation}
For instance,  we could consider  the quantum angle $A_{\Da}$ given by \eqref{qtfrhora}, but it could be any $A_f$ issued from the quantization of a function $f(\theta)$  along the procedure \eqref{qtfrhor}.

The pointer is aimed to detect the ``parallel'' orientation in the plane determined by the angle $\phi_{\parallel}$ and its orthogonal 
determined by  $\phi_{\perp}$.

The interaction pointer-system generating a measurement at time $t=t_M$ is described by  
the operator
\begin{equation}
\label{intham}
\widetilde{H}_{\mathrm{int}}(t)= \delta(t-t_M)\, \tau_2 \otimes A^S\,.
\end{equation}
Note that it is a Hamiltonian in our sense for the pointer, but it is not for the system.  However, we consider the following   corresponding  evolution operator $U(t_0,t)$ for $t_0< t_M$:
\begin{equation}
\label{nevop}
U(t,t_0)= \exp\left[ \frac{1}{\hbar}\int_{t_0}^{t}\ud t^{\prime}\, \delta(t^{\prime}-t_M)\, \tau_2 \otimes A^S\right] = \exp\left[ \frac{1}{\hbar}\Theta(t-t_M)\, \tau_2 \otimes A^S\right]\, , 
\end{equation} 
where $\Theta$ is the Heaviside function. Hence, as soon as $t>t_m$, and from the general formula involving an orthogonal projector $P$,  
\begin{equation}
\label{expP}
\exp(\lambda \tau_2 \otimes P) = \mathcal{R}(\lambda) \otimes  P + I\otimes (I-P)\, , 
\end{equation}
we obtain
\begin{equation}
\label{nevop1}
t> t_M\, , \quad U(t,t_0)= \mathcal{R}(\lambda_{\parallel}) \otimes  P_{\phi_{\parallel}} + \mathcal{R}(\lambda_{\perp}) \otimes  P_{\phi_{\perp}}\, . 
\end{equation}
One easily checks that  $U(t,t_0) U(t,t_0)^{\dag} = U(t,t_0)^{\dag} U(t,t_0)  = I\otimes I$. 
After having prepared the pointer in the state $|\phi_M=0\rg$,  the  action of \eqref{nevop1} on the initial state $|\phi_M=0\rg \otimes  |\phi_S\rg$ reads for $t> t_M$
\begin{equation}
\label{actin}
 U(t,t_0)\,|\phi_M=0\rg \otimes  |\phi_S\rg= \cos\left(\phi_S - \phi_{\parallel}\right) \, \left|\lambda_{\parallel}\right\rg\otimes\left|\phi_{\parallel}\right\rg   - \sin\left(\phi_S - \phi_{\parallel}\right) \, \left|\lambda_{\perp}\right\rg\otimes\left|\phi_{\perp}\right\rg\, . 
\end{equation}
As expected from the standard  theory of quantum measurement,  this formula indicates that the probability for the pointer to indicate the parallel orientation $\phi_{\parallel}$ is $\cos^2\left(\phi_S - \phi_{\parallel}\right)$ whereas it is $\sin^2\left(\phi_S - \phi_{\parallel}\right)$ for the perpendicular orientation $\phi_{\parallel} + \pi/2 = \phi_{\perp}$. 


\section{Conclusion}
\label{conclu}

Basic aspects of quantum formalism have been presented in this paper, taking the Euclidean plane as the simplest mathematical model to be considered in a pedagogical approach. We have shown the central role played by the unit circle on both classical and quantum levels. The unit circle is not only viewed as the set of pure states, but it is also the classical model from which are built quantum observables through a covariant integral quantization. In the presentation of the procedure, we have insisted on its probabilistic aspects. The corresponding quantum dynamics has been given a physical flavour at the price of increasing the dimension from two to three in order to implement what we call integral quantization, and a simple illustration is presented in terms of magnetic moment.  We have shown that entanglement,  a central notion of modern developments in quantum physics, allows to follow an interesting path  from the Euclidean plane to the quantum one-half spin formalism and to the quaternionic language as well. As an interesting continuation of the procedure, this path can be extended in order to include entanglement of two $1/2$ spins  and to give an interesting description of the latter  in an octonionic language \cite{dray_manogue15}. Indeed, like $\R^2 \otimes \R^2 \cong \C^2 \cong \H$ (as vector spaces), we can continue with $\C^2 \otimes \C^2 \cong \C^4 \cong \H \times \H \cong \mathbb{O}$ (as vector spaces), the Cartesian product of two quaternion fields being given an octonionic structure \textit{\`a la} Cayley Dickson.   This is the content of a  work in progress.

\appendix

\section{Parametrizations of $2\times2$ real density matrices}
\label{pararho}

There are various expressions for a density matrix acting on the Euclidean plane, i.e. a $2\times2$ real positive matrix with trace equal to $1$. The most immediate one is the following with parameters $a$ and $b$: 
\begin{equation}
\label{dens-a-b}
\rho := \mathsf{M}(a,b) = \begin{pmatrix}
   a   &  b \\
    b  &  1-a
\end{pmatrix}\, , \quad 0\leq a\leq 1\, , \quad \Delta:= \det \rho = a(1-a)-b^2 \geq 0\,. 
\end{equation}
The above inequalities imply the following ones
 \begin{equation}
\label{ineqab}
 0\leq a(1-a)\leq \frac{1}{4}\, , \quad 0\leq \Delta \leq \frac{1}{4}\, , \quad   -\frac{1}{2} \leq b \leq \frac{1}{2}\,. 
\end{equation}
Let 
\begin{equation}
\label{eigrho}
\frac{1}{2} \leq \lambda = \frac{1}{2}(1+\sqrt{1-4\Delta}) \leq 1
\end{equation}
 be the highest eigenvalue of $\rho$ (the lowest one is $0\leq 1-\lambda\leq 1/2$). The spectral decomposition of $\rho$ reads as
\begin{equation}
\label{spedec}
\rho = \lambda |\phi\rg\lg\phi| + (1-\lambda) \left|\phi + \frac{\pi}{2}\right\rg\left\lg\phi+ \frac{\pi}{2}\right| \end{equation}
where 
\begin{equation}
\label{vectphi}
|\phi\rg \equiv \begin{pmatrix}
    \cos\phi      \\
    \sin\phi    
\end{pmatrix} \, , \quad - \frac{\pi}{2}\leq \phi\leq \frac{\pi}{2}\, ,
\end{equation}
is the corresponding unit eigenvector, chosen as pointing toward the right half-plane. We could have as well chosen the opposite $|\phi + \pi\rg= -|\phi\rg$ pointing in the left half-plane since $|\phi + \pi \rg\lg\phi + \pi| = |\phi\rg\lg\phi|$. Our choice corresponds to the most immediate one in terms of orthonormal basis of the plane issued from the canonical one $\{ |0\rg\, , \, |\pi/2\rg\}$ through the rotation by $\phi$. 

Let us make explicit the decomposition \eqref{spedec}, 
\begin{equation}
\label{spedec1}
\rho = \begin{pmatrix}
  \left(\lambda-\frac{1}{2}\right)\cos(2\phi) + \frac{1}{2}    &    \left(\lambda-\frac{1}{2}\right)\sin(2\phi) \\
   \left(\lambda-\frac{1}{2}\right)\sin(2\phi)    &    \left(\frac{1}{2}- \lambda\right)\cos(2\phi) + \frac{1}{2} 
\end{pmatrix}\, . 
\end{equation}
We derive from this expression the polar parametrization of the $(a,b)$ parameters of $\rho$:
\begin{equation}
\label{polpar-a-b}
a-\frac{1}{2} =  \left(\lambda-\frac{1}{2}\right)\cos(2\phi) \, , \quad b = \left(\lambda-\frac{1}{2}\right)\sin(2\phi)\, . 
\end{equation}
In return, we have the angle $\phi\in [-\pi/2,\pi/2]$ in function of $a$ and $b$
\begin{equation}
\label{phiab}
\phi = \left\lbrace\begin{array}{cc}
 \frac{1}{2} \arctan\left(\dfrac{b}{a-1/2}\right) \, ,     &  - \dfrac{\pi}{4}\leq \phi\leq \dfrac{\pi}{4} \, ,  \\
       \frac{1}{2} \arctan\left(\dfrac{b}{a-1/2}\right)  + \dfrac{\pi}{4} \, ,   & \vert \phi\vert \geq \dfrac{\pi}{4}    \, . 
\end{array}   \right.
\end{equation}
In this way, each $\rho$ is univocally (but not biunivocally) determined by a point in the  unit disk, with polar coordinates $(r:=2\lambda-1, \Phi:=2\phi)$, $0\leq r\leq1$, $-\pi\leq \Phi <\pi$. 

Also note the alternative expression issued from \eqref{spedec1}:
\begin{equation}
\label{spedec2}
\rho = \mathsf{R}(r,\Phi)=  \frac{1}{2}(I +   r \,\mathcal{R}(\Phi) \sigma_3)= \frac{1}{2}(I +     (2\lambda-1)\,\mathcal{R}(\phi) \sigma_3\mathcal{R}(-\phi))\, ,  
\end{equation}
where $\mathcal{R}(\Phi)$ is the rotation matrix in the plane
\begin{equation}
\label{rotmatA}
\mathcal{R}(\Phi) = \begin{pmatrix}
\cos\Phi &  -\sin\Phi\\
   \sin\Phi    &    \cos\Phi
\end{pmatrix}\, ,
\end{equation}
and $\sigma_3$ is the diagonal Pauli matrix
\begin{equation}
\label{pauli3}
\sigma_3 = \begin{pmatrix}
1 & 0\\
0 &  -1
\end{pmatrix}\, .
\end{equation}
In order to get the second equality in \eqref{spedec2} we used the important property
\begin{equation}
\label{rotmatsig}
\mathcal{R}(\Phi) \sigma_3= \begin{pmatrix}
\cos\Phi &  \sin\Phi\\
   \sin\Phi    &  -  \cos\Phi
\end{pmatrix}= \sigma_3\mathcal{R}(-\Phi) \, . 
\end{equation}
Therefore the expression of a matrix density  to which we  refer mostly often through the paper 
reads
\begin{equation}
\label{standrho}
\rho = \mathsf{R}(r,\Phi) = \begin{pmatrix}
  \frac{1}{2}  + \frac{r}{2}\cos\Phi  &   \frac{r}{2}\sin\Phi  \\
\frac{r}{2}\sin\Phi     &   \frac{1}{2}  - \frac{r}{2}\cos\Phi
\end{pmatrix}\, . 
\end{equation}

The expression \eqref{standrho} is convenient to examine the way a density matrix transforms under a rotation $\mathcal{R}(\omega)$ in the plane. We have the covariance property 
\begin{align}
\label{rotrho}
\nonumber \rho&= \mathsf{R}(r,\Phi) \mapsto \mathcal{R}(\omega) \mathsf{R}(r,\Phi)\mathcal{R}(-\omega)=  \frac{1}{2}(I +     (2\lambda-1)\,\mathcal{R}(\phi + \omega) \sigma_3\mathcal{R}(-\phi - \omega))\\ &= \mathsf{R}(r,\Phi + 2\omega) \equiv \rho(\omega)\, . 
\end{align}

\section{SU$(2)$ as unit quaternions acting in $\mathbb{R}^{3}$\label{sec:su2}}
\label{allSU2}

\subsection{Rotations and quaternions }

A convenient representation of rotations in space is possible thanks to quaternion calculus.
We recall that the quaternion field as a multiplicative group is $\mathbb{H}\simeq\mathbb{R}_{+}\times$SU$(2)$.
The correspondence between the canonical basis of $\mathbb{H}\simeq\mathbb{R}^{4}$,
$(1\equiv \mathsf{\mathbf{f}}_{0},\mathsf{\mathbf{f}}_{1},\mathsf{\mathbf{f}}_{2},\mathsf{\mathbf{f}}_{3})$, and the Pauli matrices is $\mathsf{\mathbf{f}}_{a}\leftrightarrow(-1)^{a+1}\ii\sigma_{a}$,
with $a=1,2,3$. Hence, the $2\times2$ matrix representation of these
basis elements is the following: 
\begin{equation}
\label{corSU2H}
\begin{pmatrix}1 & 0\\
0 & 1
\end{pmatrix}\leftrightarrow \mathsf{\mathbf{f}}_{0}\,,\,\begin{pmatrix}0 & \ii\\
\ii & 0
\end{pmatrix}\leftrightarrow \mathsf{\mathbf{f}}_{1}\equiv\widehat{\boldsymbol{\imath}}\,,\,\begin{pmatrix}0 & -1\\
1 & 0
\end{pmatrix}\leftrightarrow \mathsf{\mathbf{f}}_{2}\equiv\widehat{\boldsymbol{\jmath}}\,,\,\begin{pmatrix}\ii& 0\\
0 & -\ii
\end{pmatrix}\leftrightarrow \mathsf{\mathbf{f}}_{3}\equiv\widehat{\mbox{\textbf{\textit{k}}}}\,,
\end{equation}
where we have also introduced  more familiar symbols for the 

Any quaternion decomposes as $q=(q_{0},\vec{\mathbf{ q}})$ (resp. $q^{a}\mathsf{\mathbf{f}}_{a},a=0,1,2,3$)
in scalar-vector notation (resp. in Euclidean metric notation). We
also recall that the multiplication law explicitly reads in scalar-vector
notation: $qq^{\prime}=(q_{0}q_{0}^{\prime}-\vec{\mathbf{q}}\cdot\vec{\mathbf{q}^{\prime}},q_{0}^{\prime}\vec{\mathbf{q}}+q_{0}\vec{\mathbf{q}^{\prime}}+\vec{\mathbf{q}}\times\vec{\mathbf{q}^{\prime}})$.
The (quaternionic) conjugate of $q=(q_{0},\vec{\mathbf{q}})$ is $\bar{q}=(q_{0},-\vec{\mathbf{q}})$,
the squared norm is $\Vert q\Vert^{2}=q\bar{q}$, and the inverse
of a nonzero quaternion is $q^{-1}=\bar{q}/\Vert q\Vert^{2}$. Unit
quaternions, i.e., quaternions with norm $1$, which form the multiplicative
subgroup isomorphic to SU$(2)$, constitute the three-sphere $\mathbb{S}^{3}$.

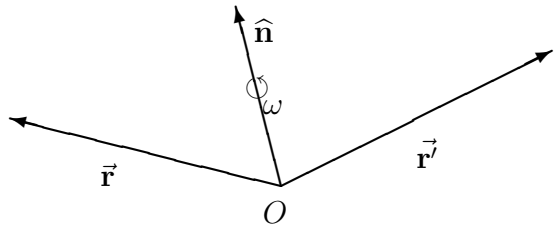
\begin{figure}
\begin{center}
 \setlength{\unitlength}{1.2mm}
\begin{picture}(40,40) 
\thicklines \put(20,10){\vector(2,1){30}} 
\put(18,6){$O$}
\put(35,12){$\vec{\mathbf{r}^{\prime}}$} 
\put(20,10){\vector(-4,1){30}} 
\put(16,20){$\circlearrowleft$}
\put(18,18){$\omega$} 
\put(17,26){$\widehat{\mathbf{n}}$} 
\put(20,10){\vector(-1,4){5}}
\put(0,10){$\vec{\mathbf{r}}$} 
\end{picture}
\caption{
The direct rotation by 
$\omega$
 about the oriented axis  $(O,\widehat{\mathbf{n}})$ brings vector $\vec{\mathbf{r}}$ to vector $\vec{\mathbf{r}^{\prime}}$. 
}
\label{rotation}
\end{center}
\end{figure}

On the other hand, any proper rotation in space is determined by a
unit vector $\widehat{\mathbf{n}}$ defining the rotation axis and a rotation angle
$0\leq\omega<2\pi$ about the axis, as illustrated in Figure \ref{rotation}.

The action of such a rotation, $\mathcal{R}(\omega,\widehat{\mathbf{n}})$,
on a vector $\vec{\mathbf{r}}$ is given by: 
\begin{equation}
\vec{\mathbf{ r}^{\prime}}\overset{def}{=}\mathcal{R}(\omega,\widehat{\mathbf{n}})\cdot\vec{\mathbf{r}}=\vec{\mathbf{r}}\cdot\widehat{\mathbf{n}}\,\widehat{\mathbf{n}}+\cos\omega\,\widehat{\mathbf{n}}\times(\vec{\mathbf{r}}\times\widehat{\mathbf{n}})+\sin\omega\,(\widehat{\mathbf{n}}\times\vec{\mathbf{r}})\,.\label{rotnot1}
\end{equation}

The latter is expressed in scalar-vector quaternionic form as 
\begin{equation}
(0,\vec{\mathbf{ r}^{\prime}})=\xi(0,\vec{\mathbf{ r}})\bar{\xi}\,,
\end{equation}
where 
\begin{equation}
\xi:=\left(\cos\frac{\omega}{2},\sin\frac{\omega}{2}\,\widehat{\mathbf{n}}\right)\in\mathrm{SU}(2)\,,
\end{equation}
or, in matrix form resulting from the identification \eqref{corSU2H},
\begin{equation}
\xi  \equiv\left(\begin{array}{cc}
\xi_{0}+\ii\xi_{3} & -\xi_{2}+\ii\xi_{1}\\
\xi_{2}+\ii\xi_{1} & \xi_{0}-\ii \xi_{3}
\end{array}\right)
  =\left(\begin{array}{cc}
\cos\frac{\omega}{2}+\ii n^{3}\sin\frac{\omega}{2} & \left(-n^{2}+\ii n^{1}\right)\sin\frac{\omega}{2}\\
\left(n^{2}+\ii n^{1}\right)\sin\frac{\omega}{2} & \cos\frac{\omega}{2}-\ii n^{3}\sin\frac{\omega}{2}
\end{array}\right)\,,\label{eq:xi-matrix}
\end{equation}
in which case quaternionic conjugation corresponds to the transposed
conjugate of the corresponding matrix,
\begin{equation}
\label{invquat}
\bar \xi= \xi^{-1}= \begin{pmatrix}
          \xi_{0}-\ii\xi_{3} & \xi_{2}-\ii\xi_{1}\\
-\xi_{2} -\ii\xi_{1} & \xi_{0}+\ii \xi_{3}
        \end{pmatrix}\,. 
\end{equation}
In particular, for a given unit vector 
\begin{align}
\label{unitvecS2}\widehat{\mathbf{n}} (\theta,\phi)& =(\sin\theta\cos\phi,\sin\theta\sin\phi,\cos\theta)=\widehat{\mathbf{n}} \,,  \\
\nonumber & 0\leq\theta\leq\pi\,,\quad0\leq\phi<2\pi\,,
\end{align}
one considers the specific rotation $\mathcal{R}(\theta,\widehat{\mathbf{u}}_{\phi})$ that
maps the unit vector pointing to the north pole, $\widehat{\mbox{\textbf{\textit{k}}}}=(0,0,1)$,
to $\widehat{\mathbf{n}}$, as is shown in Figure \ref{rotkton}, 
\begin{equation}
\left(0,\widehat{\mathbf{n}}\right)=\left(0,\mathcal{R}(\theta,\widehat{\mathbf{u}}_{\phi})\widehat{\mbox{\textbf{\textit{k}}}}\right)\equiv\xi_{\widehat{\mathbf{n}}}\left(0,\widehat{\mbox{\textbf{\textit{k}}}}\right)\bar{\xi}_{\widehat{\mathbf{n}}}\,,\quad\widehat{\mathbf{u}}_{\phi}\overset{def}{=}(-\sin\phi,\cos\phi,0)\,,\label{rotspec}
\end{equation}
with, in quaternionic \underline{and} matrix notations,  
\begin{equation}
\label{xir} 
\xi_{\widehat{\mathbf{n}}}=\left(\cos\frac{\theta}{2},\sin\frac{\theta}{2}\,\widehat{\mathbf{u}}_{\phi}\right) = \begin{pmatrix}
   \cos\frac{\theta}{2}   & - \sin\frac{\theta}{2} e^{\ii \phi}  \\
  \sin\frac{\theta}{2} e^{-\ii \phi}   &  \cos\frac{\theta}{2}
\end{pmatrix}\,, \quad \bar\xi_{\widehat{\mathbf{n}}}= \begin{pmatrix}
   \cos\frac{\theta}{2}   & \sin\frac{\theta}{2} e^{\ii \phi}  \\
  -\sin\frac{\theta}{2} e^{-\ii \phi}   &  \cos\frac{\theta}{2}
\end{pmatrix}\,,
\end{equation}
\begin{equation}
\label{nquat} (0, \widehat{\mathbf{n}})  = \widehat{\mathbf{n}} =  \begin{pmatrix}
  \ii\cos \theta   &    \ii \sin \theta e^{\ii\phi} \\
  \ii \sin \theta e^{-\ii\phi}     &  -\ii \cos\theta
\end{pmatrix}\,. 
\end{equation}

\begin{figure}[htb!]
\begin{center}
\setlength{\unitlength}{1.5mm} 
\begin{picture}(50,30)
\thicklines \put(20,10){\vector(1,1){16}} 
\put(20,10){\vector(4,1){15}}
\put(18,6){$O$} 
\put(36,22){$\widehat{\mathbf{n}}$} 
\put(33,9){$\widehat{\mathbf{u}}_{\phi}$}
\put(20,12.5){$\curvearrowright$} 
\put(27,11.5){$\curvearrowright$}
\put(22,16){$\theta$} 
\put(17,28){$\widehat{\mbox{\textbf{\textit{k}}}}$} 
\put(20,10){\vector(0,4){23}}
\end{picture}
\caption{ Specific rotation $\mathcal{R}_{\widehat{\mathbf{n}}}$ that
maps the unit vector pointing to the north pole, $\widehat{\mbox{\textbf{\textit{k}}}}=(0,0,1)$,
to $\widehat{\mathbf{n}}$.}
\label{rotkton}
\end{center}
\end{figure}

\subsection{Spin $1/2$ formalism}
The unitary irreducible representations $\xi \mapsto D^j(\xi)$  of SU$(2)$ \cite{talman68} are labelled by half-integers $j$. For a given $j$, the corresponding complex Hilbert space is finite-dimensional, with dimension $2j+1$. The matrix elements of the linear operator $D^j(\xi)$ in the basis $|j,m\rg$ are given by 
\begin{align}
\label{matelsu2apc}
\nonumber  D^j_{m_1 m_2} (\xi) &= (-1)^{m_1 - m_2} \left\lbrack (j+m_1)! (j-m_1)! (j+m_2)! (j-m_2)! \right\rbrack^{1/2} \times \\
 \times &\sum_{t} \frac{( \xi_0 + i \xi_3)^{j-m_2 - t}}{(j - m_2 -t)!} \, \frac{( \xi_0 - i \xi_3)^{j +m_1 - t}}{(j +  m_1 -t)!}\, \frac{(- \xi_2 + i \xi_1)^{t+m_2 - m_1}}{(t +m_2 - m_1)!}  \frac{( \xi_2+ i \xi_1)^{ t}}{t!} \, ,
\end{align}
in agreement with  Talman \cite{talman68}. In the lowest non trivial dimensional case we are concerned with in this paper, $j=1/2$, these matrix elements are given by
\begin{equation}
\label{mateldj12}
D^{\frac{1}{2}}_{\frac{1}{2} \frac{1}{2}}(\xi)= \xi_0 - \ii \xi_3 \, , \quad D^{\frac{1}{2}}_{-\frac{1}{2} \frac{1}{2}}(\xi)=  -\xi_2 + \ii \xi_1 \, , \quad  D^{\frac{1}{2}}_{\frac{1}{2} -\frac{1}{2}}(\xi)= \xi_2 + \ii \xi_1\, , \quad  D^{\frac{1}{2}}_{-\frac{1}{2} -\frac{1}{2}}(\xi) = \xi_0 + \ii \xi_3 \,. 
\end{equation}
Hence, we have to be careful when we work with matrix representations of quaternions, due to the choice of basis. Consequently, besides the ``canonical'' one \eqref{eq:xi-matrix}, we  use preferably (and safely!) the notation $D^{\frac{1}{2}}(\xi)$ for the $2\times 2$ matrix given by \eqref{mateldj12} in the ``up'' and ``down'' spin basis,
\begin{equation}
\label{updownnot}
\left| \frac{1}{2} \frac{1}{2}\right\rg = \upa\, , \qquad \left| \frac{1}{2}-\frac{1}{2}\right\rg = \dwa\, .
\end{equation}
Precisely, we have
\begin{align}
\label{D12xi}
  D^{\frac{1}{2}}(\xi)  & = (\xi_0 + \ii \xi_3) \dwa\wda +  (-\xi_2 + \ii \xi_1) \dwa\pua + ( \xi_2 + \ii \xi_1)\upa\wda +  (\xi_0 - \ii \xi_3) \upa\pua\, , \\
  \label{D12bxi} D^{\frac{1}{2}}(\bar \xi) & = (\xi_0 - \ii \xi_3) \dwa\wda +  (\xi_2 - \ii \xi_1) \dwa\pua + ( -\xi_2 - \ii \xi_1)\upa\wda +  (\xi_0 + \ii \xi_3) \upa\pua \, ,  
\end{align}
and for the particular case $\xi= \xi_{\widehat{\mathbf{n}}}$, essential in the construction of spin $1/2$ coherent states, 
\begin{align}
\label{D12xin}
  D^{\frac{1}{2}}\left(\xi_{\widehat{\mathbf{n}}}\right)  & = \cos \frac{\theta}{2} \dwa\wda   -\sin \frac{\theta}{2}\,e^{\ii \phi} \dwa\pua + \sin \frac{\theta}{2}\,e^{-\ii \phi} \upa\wda +  \cos \frac{\theta}{2} \upa\pua\, , \\
  \label{D12bxin} D^{\frac{1}{2}}\left(\bar \xi_{\widehat{\mathbf{n}}}\right) & = \cos \frac{\theta}{2} \dwa\wda +  \sin \frac{\theta}{2}\,e^{\ii \phi} \dwa\pua - \sin \frac{\theta}{2}\,e^{-\ii \phi} \upa\wda +  \cos \frac{\theta}{2} \upa\pua \, ,  
\end{align}

\section{Integral quantization of the unit 2-sphere}
\label{unitsphere}
We consider   the unit sphere equipped with its rotationally invariant  measure:
\begin{equation}
\label{measS2}
X= \mathbb{S}^2\, , \quad \mathrm{d}\nu(x) = \frac{\sin\theta\,\ud \theta\, \ud \phi}{2\pi}\, , \quad \theta \in [0, \pi]\, , \quad \phi\in [0,2\pi)\, .
\end{equation}
The Hilbert space is   $\mathcal{H}= \C^2$.  

The unit ball $\mathbb{B}$ in $\mathbb{R}^{3}$ parametrizes the
set of $2\times2$ complex density matrices $\rho$. Indeed, given
a 3-vector $\vec{\mathbf{r}}\in\mathbb{R}^{3}$, viewed as a pure quaternion $\vec{\mathbf{r}}\equiv (0,\vec{\mathbf{r}})\in \mathbb{H}$, and  such that $r = \Vert\vec{\mathbf{r}}\Vert\leq1$,
a general density matrix $\rho$ can be written as 
\begin{equation}
\label{densrhod}
\rho\equiv \rho_{\vec{\mathbf{r}}}= \frac{1}{2}\left(1-\ii\,D^{\frac{1}{2}}\left( \vec{\mathbf{r}}\right)\right)=  \frac{1}{2}\,\begin{pmatrix}
 1 + x_3     &x_1 +  \ii x_2   \\
 x_1 -  \ii x_2     &  1- x_3
\end{pmatrix} = \frac{1}{2}\, \begin{pmatrix}
 1  + r\cos \eta    &  r\sin\eta e^{\ii \zeta}  \\
   r\sin\eta e^{-\ii \zeta}   &  1  - r\cos \eta  
\end{pmatrix}\,, 
\end{equation}
where $(x_1,x_2,x_3)$   (resp. $(r, \eta, \zeta)$) are the cartesian (resp. spherical) coordinates of vector $\vec{\mathbf{r}}$. 
We have used the representation $D^{\frac{1}{2}}$ defined by \eqref{D12xi} in order to avoid any ambiguity about the basis in which is expressed the density matrix $\rho$. 
If $r=1$, with spherical coordinates $(\eta = \theta,\zeta= -\phi)$, then $\rho$
is the pure state 
\begin{equation}
\label{purstarho}
\rho=\left\vert \theta,\phi\right\rangle \left\langle \theta,\phi\right\vert \,,
\end{equation}
where the column vector  $\left\vert \theta,\phi\right\rangle$  is the spin $j=1/2$
coherent state introduced in \eqref{spinCS}.

Let us now transport the density matrix $\rho$
by using the two-dimensional complex representation of rotations in
space, namely the matrix SU$(2)$ representation.  For $\xi\in \mathrm{SU}\left(2\right)$, one defines
the family of density matrices labelled by $\xi$: 
\begin{equation}
\label{rhorxi}
\rho_{\vec{\mathbf{r}}}(\xi):=D^{\frac{1}{2}}(\xi)\rho D^{\frac{1}{2}}(\bar{\xi})=\frac{1}{2}\left(1-\ii D^{\frac{1}{2}}\left(\xi\vec{\mathbf{r}}\bar{\xi}\right)\right)\,.
\end{equation}
In order to get a one-to-one correspondence with the points of the 2-sphere, we restrict the elements of SU$(2)$ to 
those corresponding to the rotation $\mathcal{R}(\theta,-\widehat{\mathbf{u}}_{\phi})$ bringing the unit vector $\widehat{\mbox{\textbf{\textit{k}}}}$ pointing to the North pole  to the vector $\widehat{\mathbf{n}}_-$ with spherical coordinates $(\theta,-\phi)$, as described in \eqref{rotspec} and \eqref{xir} but with an opposed sign for $\phi$ in  order to get  \eqref{purstarho} for $r=1$.
\begin{equation}
\label{rotspecmain}
\rho_{\vec{\mathbf{r}}}(\theta,\phi):= D^{\frac{1}{2}}\left(\xi_{\widehat{\mathbf{n}}_-}\right)\,\rho_{\vec{\mathbf{r}}}\,D^{\frac{1}{2}}\left(\bar \xi_{\widehat{\mathbf{n}}_-}\right)= \rho_{ \xi_{\widehat{\mathbf{n}}_-}\vec{\mathbf{r}}\bar \xi_{\widehat{\mathbf{n}}_-}}\, , 
\end{equation}
The value of the integral 
\begin{equation}
\int_{\mathbb{S}^2} \rho_{\vec{\mathbf{r}}}(\theta,\phi)\,\frac{\sin\theta\,\ud \theta\, \ud \phi}{2\pi}= \begin{pmatrix}
  1    & \dfrac{x_1+\ii x_2}{2}   \\
 \dfrac{x_1-\ii x_2}{2}     &  1
\end{pmatrix} \label{resmatS2}
\end{equation}
shows that the resolution of the unity is achieved with $\vec{\mathbf{r}}=\pm r\,\widehat{\mbox{\textbf{\textit{k}}}}$ only. Then, it is clear that 
\begin{equation}
\label{dkrotr}
\rho_{r \widehat{\mbox{\textbf{\textit{k}}}} }(\theta,\phi) =  \frac{1}{2}\begin{pmatrix}
1+ r\cos\theta      &  r\sin\theta\,e^{-\ii\phi}  \\
   r\sin\theta\,e^{\ii\phi}     &  1- r\cos\theta 
\end{pmatrix}\,.  
\end{equation}
It is with this strong restriction and the simplified notation 
\begin{equation}
\label{simplnot}
 \rho_{r\widehat{\mbox{\textbf{\textit{k}}}}}(\theta,\phi) \equiv \rho_r(\theta,\phi)
\end{equation}
that we go forward to the next calculations with the resolution of the unity 
\begin{equation}
\label{S2resun}
\int_{\mathbb{S}^2} \rho_{r}(\theta,\phi)\,\frac{\sin\theta\,\ud \theta\, \ud \phi}{2\pi}= I\,. 
\end{equation} 
Note that  the resolution of the identity with the SU$(2)$ transport of  a generic density operator \eqref{densrhod} is possible only if we integrate on the whole group, as it was done in \cite{balfrega14}.

The $\mathbb{S}^2$-labelled family of probability distributions on $(\mathbb{S}^2\,, \, \sin\theta\,\ud \theta\, \ud \phi/2\pi)$
\begin{align}
\label{probdisph}
 \nonumber p_{\theta_0,\phi_0}(\theta,\phi) &= \mathrm{tr}\left(\mathsf{\rho_{r}}(\theta_0,\phi_0)\,\mathsf{\rho_{r}}(\theta,\phi)\right)= \frac{1}{2}\left(1 + r^2 \widehat{\mathbf{r}}_0\cdot \widehat{\mathbf{r}})\right)\\
 &= \frac{1}{2}\left(1 + r^2(\cos\theta_0\cos\theta + \sin\theta_0\sin\theta\cos(\phi_0-\phi) \right)\, . 
\end{align}
At $r= 0$ we get the uniform probability on the sphere whereas at $r=1$ we get the probability distribution corresponding to the spin 1/2 CS \eqref{spinCS},
\begin{equation}
\label{purdistsph}
p_{\theta_0,\phi_0}(\theta,\phi) = \vert \lg \theta_0,\phi_0 | \theta,\phi\rg \vert^2 \, . 
\end{equation}


The quantization of a function (or distribution) $f(\theta,\phi)$ on the sphere based on \eqref{S2resun}   leads to 
the 2$\times$2 matrix operator
\begin{equation}
\label{qtfrhorS2}
f \mapsto A_f = \int_{\mathbb{S}^2} f(\theta,\phi) \rho_{r}(\theta,\phi) \,\frac{\sin\theta\,\ud \theta\, \ud \phi}{2\pi} = \begin{pmatrix}
  \lg f\rg + r\,C^{\mathbb{S}^2}_c (f)  &   r\,C^{\mathbb{S}^2}_s(f) \\
r\,\left(C^{\mathbb{S}^2}_s (f )\right)^{\ast}   &  \lg f \rg  - r\,C^{\mathbb{S}^2}_c(f)
\end{pmatrix}\,,
\end{equation}
where $ \lg f\rg:= \frac{1}{4\pi}\int_{\mathbb{S}^2}f(\theta,\phi)\,\sin\theta\,\ud \theta\, \ud \phi$ is the average of $f$ on the unit sphere and  $C^{\mathbb{S}^2}_c$ and $C^{\mathbb{S}^2}_s$ are  Fourier coefficients of $f$ on the sphere defined as
\begin{equation}
\label{CcCsS2}
C^{\mathbb{S}^2}_c(f) = \frac{1}{4\pi}\int_{\mathbb{S}^2}f(\theta,\phi)\,\cos\theta\,\sin\theta\,\ud \theta\, \ud \phi\, , \quad C^{\mathbb{S}^2}_s(f) = \frac{1}{4\pi}\int_{\mathbb{S}^2}f(\theta,\phi)\,e^{-\ii \phi}\, \sin^2\theta\,\ud \theta\, \ud \phi\, . 
\end{equation}
Applying the quantization formula \eqref{qtfrhorS2} of the three components of the generic  unit vector \eqref{unitvecS2}  provides, as expected, the three Pauli matrices, up to a common factor in agreement with general results given in \cite{gahulare07}, 
\begin{equation}
\label{ntosig}
\widehat{\mathbf{n}} \mapsto A_{\widehat{\mathbf{n}}} = \int_{\mathbb{S}^2} \begin{pmatrix}
      \sin\theta \cos\phi\\
      \sin\theta \sin\phi\\
      \cos \theta  
\end{pmatrix}  \rho_{r}(\theta,\phi) \,\frac{\sin\theta\,\ud \theta\, \ud \phi}{2\pi} = \frac{r}{3} \begin{pmatrix}
      \sigma_1    \\
      \sigma_2 \\
      \sigma_3  
\end{pmatrix}\,. 
\end{equation}

\subsection*{Acknowledgments}
J.-P. Gazeau thanks CBPF and  CNPq for financial support and CBPF for hospitality. E.M.F. Curado acknowledges CNPq and FAPERJ for financial support. The authors are grateful to  Fernando  de Melo (CBPF), Pierre Martin-Dussaud (ENS Lyon) and  Tomoi Koide (UFRJ, Rio) for helpful comments.


\newpage

\begin{thebibliography}{99}



\bibitem{euclid300BC}   Euclid's Elements,  a bilingual edition in English and Greek, Ed. \& trans. R. Fitzpatrick, utexas.edu, circa 300 BC; pdf available at \url{http://farside.ph.utexas.edu/Books/Euclid/Elements.pdf}  


\bibitem{stueck60} E.C.G. Stueckelberg, Quantum theory in real Hilbert space, \textit{ Helv. Phys. Acta} \textbf{33} 727-752 (1960). 

\bibitem{gazeaubook09} J.-P. Gazeau,  
\textit{Coherent States in Quantum Physics},  Wiley-VCH, Berlin, 2009.

\bibitem{aagbook13} S.~T. Ali, J.-P Antoine, and J.-P. Gazeau, \textit{Coherent
States, Wavelets and their Generalizations} 2d edition, Theoretical
and Mathematical Physics, Springer, New York (2013), specially Chapter
11.

\bibitem{gahell15} J.-P. Gazeau and B. Heller, Positive-Operator
Valued Measure (POVM) quantization, \textit{Axioms} (Special issue
on \emph{Quantum Statistical Inference}) \textbf{4} 1 (2015);
http://www.mdpi.com/2075-1680/4/1/1.

\bibitem{dirac39} P.~A.~M. Dirac, A new notation for quantum mechanics, \textit{Math. Proc.  Cambridge Phil. Soc.} \textbf{35}  416-418 (1939).

\bibitem{jordan} A Jordan algebra is an (not necessarily associative) algebra over a field whose multiplication is commutative and satisfies the \emph{Jordan identity}:
$(xy)x^2 = x(yx^2)$.

\bibitem{VNentropy} For a quantum-mechanical system described by a density matrix $\rho= \sum_n p_n \, |\psi_n\rg\lg \psi_n|$, the von Neumann entropy is 
$S_{\mathrm{VN}} (\rho)=- \mathrm{Tr}\rho\ln \rho = -\sum_n p_n\ln p_n$, the last expression being actually its Shannon form.  It is zero if and only if $\rho$ is a pure state, i.e., a projector. 


\bibitem{bengtsson07} I. Bengtsson  and K. Zyczkowski, \textit{Geometry of Quantum States: An Introduction to Quantum Entanglement} (1st ed.), Cambridge University Press (2007) p. 301.

\bibitem{dirac30} P.~A.~M. Dirac, \textit{The Principles of Quantum Mechanics}, 4th edition, Clarendon Press, Oxford (1981).

\bibitem{vonneumann32} J. Von Neumann, \textit{Mathematische Grundlagen der Quantenmechanik}, Springer, Berlin (1932). English translation by R.~T. Beyer, \textit{Mathematical Foundations of Quantum Mechanics}, Princeton University Press (1955).

\bibitem{omnes94} R. Omn\`es, \textit{The Interpretation of Quantum Mechanics}, Princeton (1994).

\bibitem{laloe2012} F. Lalo\"{e}, \textit{Do we really Understand Quantum Mechanics}, Cambridge U. P. (2012).

\bibitem{bricmont16} J. Bricmont, \textit{Making Sense of Quantum Mechanics}, Springer (2016).

\bibitem{zachos05} C.~K. Zachos, D.~B. Fairlie, and T.~L. Curtright, \textit{Quantum Mechanics in Phase Space}, World Scientific, Singapore (2005).


\bibitem{lieb73} E.~H. Lieb, The classical limit of quantum spin systems,
 \textit{Commun. Math. Phys.} \textbf{31}  327-340 (1973).
 
 \bibitem{berezin75} F.~A. Berezin,  Quantization,   \textit{Math. USSR Izvestija} \textbf{8}  1109-1165 (1974);
  General concept of quantization, \textit{Commun. Math. Phys.} \textbf{40}  153-174 (1975). 
  
  \bibitem{barracz77} A.~O. Barut and R. R\c{a}czka, \textit{Theory of Group Representations
       and Applications}, PWN, Warszawa, 1977.


\bibitem{mardia72} K.~V Mardia,  \textit{Statistics of Directional Data}, Academic Press, New York,  1972.

\bibitem{borelset} A Borel set is any set in a topological space that can be formed from open sets (or, equivalently, from closed sets) through the operations of countable union, countable intersection, and relative complement. 

\bibitem{sigmaalg} A $\sigma$-algebra is a  collection  of subsets of of a set that includes the empty subset, is closed under complement, and is closed under union or intersection of countably infinite many subsets.

\bibitem{delt} R. Deltheil, 
      \textit{Probabilit\'es g\'eom\'etriques}, Trait\'e de Calcul des Probabilit\'es et de ses
Applications par \'Emile Borel, Tome II, Gauthiers-Villars, Paris, 1926.



\bibitem{feynman3} R.~P. Feynman, R.~B. Leighton, M. Sands, \textit{The Feynman Lectures on Physics}, Vol. 3,  Addison-Wiley, 1964.

\bibitem{wilczek09} F. Wilczek, Majorana returns, \textit{Nature Physics} \textbf{5} 614-618 (2009). 

\bibitem{breupetr02} H.-P. Breuer and F. Petruccione, \textit{The Theory of open quantum systems}, Oxford Univ Press, 2002.

\bibitem{weinberg17} S. Weinberg, The Trouble with Quantum Mechanics, \textit{The New York Review of Books},   January 19, 2017 Issue. 
2017

\bibitem{weinberg98} N.  Wheeler, Remarks concerning the status \& some ramifications of Ehrenfest theorem, Reed College Physics Department March 1998, preprint available at \url{http://www.reed.edu/physics/faculty/wheeler/documents/Quantum\%20Mechanics/Miscellaneous\%20Essays/Ehrenfest\%27s\%20Theorem.pdf}

\bibitem{tensor} See the answer to the question \textit{What is the difference between Cartesian and Tensor product of two vector spaces?} at  \url{http://math.stackexchange.com/questions/710481/what-is-the-difference-between-cartesian-and-tensor-product-of-two-vector-spaces}

\bibitem{nielsen-chuang00} M.~A. Nielsen and I.~L.  Chuang, \textit{Quantum computation and quantum information}, Cambridge University Press 2000.

\bibitem{rieffel_polak11} E. Rieffel and W. Polak, Quantum Computing,  A Gentle Introduction, The MIT Press,  
Cambridge, Massachusetts,  London, England (2011).


\bibitem{caves_etal01} C. M. Caves, C. A. Fuchs, and P.  Rungta, Entanglement of Formation of an Arbitrary State of Two Rebits, \textit{Found. Phys. Lett.} \textbf{14} 199-212 (2001). 


\bibitem{bell64} J.~S. Bell, On the Einstein Podolski Rosen Paradox, \textit{Physics} \textbf{1} 195-200 (1964).

\bibitem{epr35} A. Einstein, N. Rosen and B. Podolsky, \textit{Phys. Rev.} \textbf{47} 777-780 (1935).

\bibitem{bohm52} D. Bohm, A suggested interpretation of the Quantum Theory in Terms of ``Hidden'' Variables, I, \textit{Phys. Rev.} \textbf{85} 166-179 (1952); idem, II,  \textit{Phys. Rev.} \textbf{85} 180-193 (1952).

\bibitem{bohm_ah57} D. Bohm and Y. Aharonov, Discussion of Experimental Proof for the Paradox of Einstein, Rosen, and Podolsky,  \textit{Phys. Rev.} \textbf{108} 1070-1076 (1957).

\bibitem{harrspek10}  N. Harrigan and R.~W. Spekkens, Einstein, Incompleteness, and the Epistemic View of Quantum States,  \textit{Found. Phys.} (2010) \textbf{40} 125-157 (2010). 

\bibitem{feynman2} R.~P. Feynman, R.~B. Leighton, M. Sands, \textit{The Feynman Lectures on Physics}, Vol. 2, Addison-Wiley 1964.

\bibitem{resnick_eisberg85} R. Resnick and R. Eisberg, \textit{Quantum Physics of Atoms, Molecules, Solids, Nuclei and Particles} (2nd ed.), John Wiley \& Sons.  1985, p. 274.

\bibitem{peres90} A. Peres, Neumark's Theorem and Quantum Inseparability, \textit{Found. Phys.} \textbf{20} 1441-1453 (1990).

\bibitem{dray_manogue15} T. Dray and C. Manogue, \textit{The geometry of the octonions}, World Scientific, Singapore (2015). 

\bibitem{talman68} J.~D. Talman,
\textit{Special Functions, A Group Theoretical Approach}, 
W.A. Benjamin, New York, Amsterdam, 1968.

\bibitem{balfrega14} M. Baldiotti, R. Fresneda, and J.-P.  Gazeau,  Three examples of covariant integral quantization,
\emph{Proceedings of Science}  \emph{ICMP 2013} 003 (2014).

\bibitem{gahulare07} J.-P. Gazeau, E. Huguet, M. Lachi\`eze-Rey,  and J. Renaud, Fuzzy spheres from inequivalent coherent states quantizations, \textit{J. Phys. A: Math. Theor.} \textbf{40} 10225-10249 (2007).

 
\end{thebibliography}
\end{document}